\documentclass[aps,prb,amsmath, amssymb, english, twocolumn,reprint,floatfix, superscriptaddress]{revtex4-1}

\usepackage{bbm}
\usepackage{comment}
\usepackage{array}                                                                          
\usepackage{hhline}
\usepackage{tikz}
\usepackage{pgfplots}
\usepackage{microtype}
\usepackage{dsfont}
\usepackage{epstopdf}
\usepackage{epsfig} 
\usepackage[applemac]{inputenx}     
\usepackage{amsmath}
\usepackage{hyperref}

\pgfplotsset{compat=1.15}

\hypersetup{
    colorlinks,
    citecolor=blue,
    filecolor=blue,
    linkcolor=blue,
    urlcolor=blue
}

\newcommand{\bq}{\mathbf{q}}

\newcommand{\bk}{\mathbf{k}}

\allowdisplaybreaks

\graphicspath{{./Images/}}

\begin{document}
\title{Fluctuations analysis of the spin susceptibility: N\'eel ordering revisited in dynamical mean field theory}
\author{Lorenzo Del Re} 
\affiliation{Department of Physics, Georgetown University, 37th and O Sts., NW, Washington,
DC 20057, USA}
\author{Georg Rohringer}
\affiliation{Institute of Theoretical Physic, University of Hamburg, 20355 Hamburg, Germany}
\affiliation{Russian Quantum Center, Skolkovo IC, Bolshoy Bulvar 30, bld. 1, Moscow, Russia 121205}

\date{\today} 


\begin{abstract}
In this paper, we revisit the antiferromagnetic (AF) phase diagram of the single-band three-dimensional half-filled Hubbard model on a simple cubic lattice studied within the dynamical mean field theory (DMFT). Although this problem has been investigated extensively in the literature, a comprehensive understanding of the impact of the different one- and, in particular, two-particle local correlation functions of DMFT on the AF transition temperature is still missing. We have, hence, performed a fluctuation analysis of $T_N$ with respect to different local bosonic fluctuations (charge, spin, particle-particle) contained in the two-particle vertex of DMFT. Our results indicate that, beyond weak coupling, the screening of the DMFT vertex by local fluctuations leads to an enhancement of $T_N$ with respect to a random phase approximation (RPA) like calculation where this vertex is replaced by the bare interaction. The overall suppression of $T_N$ in DMFT with respect to RPA is then solely due to the incoherence introduced by the DMFT self-energy in the one-particle Green's functions. This illustrates the Janus-faced role of the local moment formation in the DMFT solution of the Hubbard model, which leads to completely opposite effects in the one- and two-particle correlation functions.

\end{abstract}

\maketitle

\section{Introduction}
\label{sec:Intro}

Transitions between different states of matter accompany mankind since the earliest time of human culture. The technological utilization of phase transitions, such as the boiling of water or the melting of metals, has often triggered substantial progress in the development of societies and the quality of life. This ascertainment still holds today where, e.g., a transition from an insulating to a conducting state is of enormous technological interest. In particular, strongly correlated materials often feature a great variety of such fascinating phenomena such as correlation-driven metal-to-insulator transitions~\cite{Mott1968}, high-temperature superconductivity~\cite{Bednorz1986}, or antiferromagnetism~\cite{Neel1936}.

Unfortunately, strongly correlated electron systems are notoriously difficult to tackle theoretically. The strong interaction between the particles prevents any quasi-independent particle description such as density functional theory\cite{Kohn1965} and the exponential growth of the Hilbert space with the particle number restricts an exact diagonalization of the corresponding Hamiltonian to small systems sizes. In this respect, the dynamical mean field theory (DMFT)\cite{Georges1996} has presented a big step forward in the understanding of strongly interacting many-body systems. This approach includes {\em all purely local} correlations between the particles by means of a local frequency-dependent self-energy $\Sigma(\nu)$ and vertex $\Gamma^{\nu\nu'\omega}$. DMFT can be applied to systems with a purely local (Hubbard) interaction between the particles with a limited number of orbitals per lattice site. A good example is the single-band Hubbard model\cite{Hubbard1963}, which represents one of the most basic model Hamiltonians for correlated many-particle systems.

In spite of its simplicity, the Hubbard model features---apart from the paramagnetic state at high temperatures---a variety of interesting phases such as ferromagnetism, superconductivity\cite{Astretsov2020} or stripe orders\cite{Zheng2016,Quin2020}. At half filling, the celebrated Mott metal-to-insulator transition\cite{Mott1968} occurs at a critical value of the Hubbard interaction $U$ beyond which stable local moments appear in the system. In three (or higher) dimensions, this first-order phase transition is covered by an antiferromagnetically (AF) ordered phase, which sets in at the N\'eel temperature $T_N$. At weak coupling, this second-order phase transition is qualitatively well described by the random phase approximation while in the limit of $U\rightarrow\infty$ the Hubbard model can be mapped onto an effective Heisenberg model\cite{Schrieffer1966} with an exchange interaction $J\!\propto\! \frac{t^2}{U}$. Between these two limiting cases, DMFT provides a reasonable estimate for the transition temperature.

In spite of these successes, it has not been investigated so far {\em how} the ingredients of DMFT, the self-energy $\Sigma(\nu)$ and the vertex function $\Gamma^{\nu\nu'\omega}$, affect the N\'eel temperature. In general, it is commonly accepted that the local correlations of DMFT and the gradual emergence of a local moment reduce $T_N$ with respect to corresponding RPA calculations. This reduction has often been attributed to the screening of the static bare interaction  $U$ by the frequency dpendent vertex $\Gamma^{\nu\nu'\omega}$. While this is indeed true at (very) weak coupling\cite{Tahvildar-Zadeh1997}, the situation is not so clear at larger values of $U$. In particular, the lack of a detailed understanding of the vertex has prevented deeper insights into the role of this two-particle correlation function for the antiferromagnetic phase transition in the Hubbard model. This situation has changed in the last decade where a lot of the properties of two-particle vertex functions have been uncovered\cite{Rohringer2012}. More specifically, it has been revealed how bosonic fluctuations determine important features of the fermionic vertex\cite{Rohringer2013a,Gunnarsson2015,Tagliavini2018,Wentzell2020} and how these fluctuations are coupled to the fermionic degrees of freedom\cite{Krien2019,Stepanov2019,Krien2020a}. This deeper understanding has, for instance, allowed to quantify the impact of two-particle fluctuations in different channels on the one-particle spectral function by means of a ``Fluctuation diagnostics''\cite{Gunnarsson2015,Wu2017,Krien2020a,Schaefer2021} or a parquet decomposition approach\cite{Gunnarsson2017,Gunnarsson2018,Rohringer2018b,Schaefer2021}.
In this paper, we extend the idea of the parquet decomposition of the self-energy\cite{Gunnarsson2018} to the study of phase transitions.  In particular, we investigate the contributions of different local bosonic fluctuations in the DMFT vertex $\Gamma^{\nu\nu'\omega}$ on the transition temperature $T_N$. To this end we first approximate this vertex by just the leading bosonic fluctuations and in a second step we perform a complementary analysis where we sequentially subtract the various local fluctuations from $\Gamma^{\nu\nu'\omega}$. Our findings show that in the intermediate-to-strong-coupling regime the frequency dependence of the vertex actually leads to an enhancement of $T_N$ with respect to the bare interaction $U$ due to the increase of local spin fluctuations. Hence, the suppression of $T_N$ with respect to the RPA originates solely from the incoherence introduced in the one-particle Green's function via the self-energy $\Sigma(\nu)$. This shows the Janus-faced role of the local-moment physics of DMFT in one- and two-particle correlation functions.

Our paper is organized as follows: In Sec.~\ref{sec:Formalism} we introduce the Hubbard model and the antiferromagnetic spin susceptibility within DMFT and review the formalism for dissecting the vertex function into different bosonic fluctuations. In Sec.~\ref{sec:numerics} we present and discuss our numerical data for the fluctuation analysis of $T_N$. In Sec.~\ref{sec:spinsusc}, we discuss the momentum dependence of the (static) lattice susceptibility and Sec.~\ref{sec:conclusions} is devoted to conclusions and an outlook.

\section{Model and formalism}
\label{sec:Formalism}

\subsection{The Hubbard model and DMFT}
The analytic technique, which we put forward in this paper, can be, in principle, applied to a variety of model systems, which feature fluctuations in different scattering channels. Here, we demonstrate its validity and applicability for the three-dimensional ($3d$) single-band Hubbard model on a simple cubic lattice with nearest-neighbor hopping

\begin{equation}
    \label{equ:defhubbard}
    \hat{H}=-t_\text{3d}\sum_{\langle ij \rangle, \sigma} \hat{c}^\dagger_{i\sigma}\hat{c}_{j\sigma}+U\sum_i\hat{n}_{i\uparrow}\hat{n}_{i\downarrow}-\mu\sum_{i\sigma}\hat{n}_{i\sigma},
\end{equation}
where $\hat{c}^{(\dagger)}_{i\sigma}$ annihilates (creates) an electron with spin $\sigma\!=\!\uparrow\!,\!\downarrow$ at the lattice site $\mathbf{R}_i$ while $\hat{n}_{i\sigma}\!=\!\hat{c}^\dagger_{\sigma}\hat{c}_{i\sigma}$ corresponds to the particle number operator. Moreover, $t_\text{3d}$ denotes the amplitude for electrons hopping between nearest-neighbor lattice sites in the $3d$ lattice, $\mu$ is the chemical potential, and $U$ corresponds to the local repulsive Hubbard interaction between the particles. In the following, we use $D\!=\!2\sqrt{6}t_\text{3d}\!=\!1$ as unit of energy, which corresponds to twice the second moment of the density of states of the non-interacting system. The average number of particles per lattice site is fixed to $n\!=\!\langle\hat{n}_\uparrow\!+\!\hat{n}_\downarrow\rangle\!=\!1$.

We apply the DMFT to solve the Hubbard model in Eq.~(\ref{equ:defhubbard}). Within this approach, the electronic propagator is dressed with a local self-energy $\Sigma(\nu)$ that depends only on the Matsubara frequencies. The lattice Green's function is then obtained via Dyson's equation
\begin{equation}\label{equ:DMFTapproximation}
    G(\nu,\bk) = \frac{1}{i\nu +\mu-\epsilon_\bk-\Sigma(\nu)},
\end{equation}
where $\epsilon_\mathbf{k}\!=\!-t_\text{3d}\sum_{<0j>}e^{i\mathbf{k}\mathbf{R}_j}$ is the bare dispersion relation and the sum over $j$ runs over the six nearest neighbors of the lattice site at the origin. $\nu\!=\!\frac{\pi}{\beta}(2n\!+\!1)$, $n\in\mathds{Z}$, denotes a fermionic Matsubara frequency with $\beta\!=\!\frac{1}{T}$ being the inverse temperature. Later, we will also consider bosonic Matsubara frequencies $\omega\!=\!\frac{\pi}{\beta}2m$, $m\in\mathds{Z}$. 

The local self-energy $\Sigma(\nu)$ is obtained from an auxiliary Anderson impurity model (AIM):
\begin{equation}
    \label{equ:defsigmaloc}
    \Sigma(\nu)=i\nu+\mu-\Delta(\nu)-\frac{1}{G_\text{loc}(\nu)},
\end{equation}
where $\Delta(\nu)$ describes the hybridization between the impurity and the non-interacting bath sites and the local impurity Green's function $G_\text{loc}(\nu)$ is obtained for a given $\Delta(\nu)$ from an impurity solver such as exact diagonalization (ED) or quantum Monte Carlo (QMC). Within DMFT, the hybridization function $\Delta(\nu)$ is determined by the requirement that the local (i.e., momentum-summed) part of the lattice Green's function is equal to the local Green's function of the auxiliary AIM:
\begin{equation}
    \label{equ:selfconst}
    \sum_\mathbf{k}\underset{G(\nu,\mathbf{k})}{\underbrace{\frac{1}{i\nu +\mu-\epsilon_\bk-\Sigma(\nu)}}}=\underset{G_\text{loc}(\nu)}{\underbrace{\frac{1}{i\nu+\mu-\Delta(\nu)-\Sigma(\nu)}}},
\end{equation}
where $\sum_\mathbf{k}$ corresponds to a (normalized) integral over the first Brillouin zone. When a QMC solver is used, $G_\text{loc}(\nu)$ can be directly obtained from $\Delta(\nu)$ while for exact diagonalization the hybridization function has to be fitted to a finite number of bath parameters:
\begin{equation}
\label{equ:defdeltaED}
\Delta(\nu)=\sum_{\ell=1}^N\frac{V_\ell^2}{i\nu-\varepsilon_\ell},
\end{equation}
where $\varepsilon_\ell$ defines the on-site energy of the bath site and $V_l$ the hopping amplitude between the bath site and the impurity. Within our ED implementation, these parameters are fitted to a given $\Delta(\nu)$, which is obtained from the self-consistency relation in Eq.~(\ref{equ:selfconst}), for a finite number $N$ of bath sites via a conjugate gradient method.

\subsection{Vertex functions and susceptibilities}
\label{sec:SpinSusc}

While the single-particle Green's function $G(\nu,\mathbf{k})$ provides information about the one-particle spectral properties of the system, two-particle correlation functions are required to describe collective (bosonic) excitations of the electrons. Theoretically, such excitations can be characterized by response functions, the (physical) susceptibilities
\begin{equation}
\label{equ:defsusc}
\chi_{r}(\omega,\mathbf{q}) = \sum_ie^{-i\mathbf{q}\cdot\mathbf{R}_i}\int_0^\beta d\tau \,e^{i\omega\tau} \, \langle \hat{O}_i^r(\tau)\hat{O}_0^r(0)\rangle, 
\end{equation}
where, for the Hubbard model, we consider the charge ($r$=ch) density $\hat{O}_i^r\!=\!\hat{n}_i\!-\!n\!=\!\hat{n}_{i\uparrow}\!+\!\hat{n}_{i\downarrow}\!-\!n$, the spin ($r$=sp) density $\hat{O}_i^r\!=\!\hat{S}^z_i\!=\!\hat{n}_{i\uparrow}\!-\!\hat{n}_{i\downarrow}$ and the particle-particle ($r$=pp) pair density $\hat{O}_i^r\!=\!\hat{c}^\dagger_{i\uparrow}\hat{c}^\dagger_{i\downarrow}\!+\!\hat{c}_{i\downarrow}\hat{c}_{i\uparrow}$ at the lattice site $\mathbf{R}_i$. Completely analogous definitions hold for the AIM where the spatial index $i$ as well as the Fourier transform $\sum_i e^{-i\mathbf{q}\cdot\mathbf{R}_i}$ have to be omitted.

A (second-order) phase transition to a spatially ordered state in one of the three fluctuation channels is signaled by a divergence of the corresponding susceptibility at $\omega\!=\!0$ and $\mathbf{q}\!=\!\mathbf{q}_0$ where $\mathbf{q}_0$ defines the spatial structure of the order. In the half filled Hubbard model, the leading instability is found in the spin channel at $\mathbf{q}_0\!=\!(\pi,\pi,\pi)\!\equiv\!\boldsymbol{\Pi}$, which corresponds to antiferromagnetic spin fluctuations, i.e.,
\begin{equation}
\label{equ:defafspinsusc}
\chi_{\text{AF}}(T)=\chi_{\text{sp}}(\omega=0,\mathbf{q}=\boldsymbol{\Pi}),\quad \boldsymbol{\Pi}=(\pi,\pi,\pi).
\end{equation}
At the N\'eel temperature $T\!=\!T_N$, at which the transition to the antiferromagnetically ordered state occurs, this response function diverges, i.e.,  
\begin{equation}
    \label{equ:defTNeel}
    \chi_\text{AF}(T\to T_N)\rightarrow \infty
\end{equation}
This criterion has been used for the determination of $T_N$ in this paper.

The physical susceptibilities defined in Eq.~(\ref{equ:defsusc}) can be obtained from more general objects, the {\em generalized} susceptibilities\footnote{Here, we have already considered the case of DMFT where the generalized lattice susceptibility depends only a single momentum $\mathbf{q}$. In the most general case, this object exhibits also a dependence on two additional momenta $\mathbf{k}$ and $\mathbf{k}'$.} $\chi_{r,\mathbf{q}}^{\nu\nu'\omega}$, by summing the latter over the fermionic Matsubara frequencies $\nu$ and $\nu'$:
\begin{equation}
\label{equ:sumgeneralizedsusc}
\chi_{r}(\omega,\mathbf{q})=\frac{1}{\beta^2}\sum_{\nu\nu'}\chi_{r,\mathbf{q}}^{\nu\nu'\omega}.
\end{equation}
The generalized susceptibilities are calculated via a Bethe-Salpeter equation (BSE) from an irreducible vertex function $\Gamma_{r}^{\nu\nu'\omega}$, which is local but frequency dependent within DMFT\cite{Georges1996}. In the channel of interest, i.e., the spin channel, the relevant BSE reads
\begin{align}
\label{equ:BS}
\chi_{\text{sp},\mathbf{q}}^{\nu\nu'\omega}=&\chi_{0,\mathbf{q}}^{\nu\nu'\omega}-\frac{1}{\beta^2}\sum_{\nu_1\nu_2}\chi_{0,\mathbf{q}}^{\nu\nu_1\omega}\Gamma_{\text{sp}}^{\nu_1\nu_2\omega}\chi_{\text{sp},\mathbf{q}}^{\nu_2\nu'\omega}.
\end{align}
Here, $\chi_{0,\mathbf{q}}^{\nu\nu'\omega}\!=\!-\beta\sum_{\mathbf{k}}G(\nu,\mathbf{k})G(\nu\!+\!\omega,\mathbf{k}\!+\!\mathbf{q})\delta_{\nu\nu'}$ is the bare susceptibility (``bubble'') of DMFT. The local irreducible vertex in the spin channel $\Gamma_{\text{sp}}^{\nu\nu'\omega}$, on the other hand, can be calculated via a purely local Bethe-Salpeter equation [analogous to Eq.~(\ref{equ:BS})] from the corresponding local generalized susceptibility $\chi_{\text{sp}}^{\nu\nu'\omega}$, which, in turn, is obtained directly from the AIM related to the DMFT solution of the Hubbard model in Eq.~(\ref{equ:defhubbard}) by means of the impurity solver (see, e.g., Ref.~\onlinecite{Toschi2007a}).

\subsection{Fluctuation decomposition of \texorpdfstring{$\Gamma_\text{sp}^{\nu\nu'\omega}$}{GammaSpin}}

\label{sec:directdiag}

The main goal of our paper is to understand how the frequency-dependent vertex $\Gamma_\text{sp}^{\nu\nu'\omega}$ affects the generalized and the physical susceptiblities in Eqs.~(\ref{equ:BS}) and (\ref{equ:sumgeneralizedsusc}), respectively. More specifically, we want to understand {\sl how} the frequency dependence of $\Gamma_\text{sp}^{\nu\nu'\omega}$ changes the related $T_N$ with respect to a simple random phase like calculation where the irreducible vertex in Eq.~(\ref{equ:BS}) is replaced by the bare interaction $U$. Moreover, we aim at unraveling {\sl which} (Feynman diagrammatic) contributions to $\Gamma_\text{sp}^{\nu\nu'\omega}$ lead to specific modifications of the critical temperature, resulting in an increase or a reduction of $T_N$. To make such an analysis meaningful, it would be of course highly desirable to identify the parts of the vertex to which a transparent physical meaning can be assigned.

In the last decade, considerable progress has been made in the understanding of the frequency structure of two-particle vertex functions\cite{Kunes2011,Rohringer2012,Hafermann2014,Rohringer2018a,Krien2019,Stepanov2019,Chalupa2018,Springer2020,Wentzell2020,chalupa2021}. In particular, it has been demonstrated that some of the main features of these correlation functions (which are also responsible for their high-frequency behavior) correspond to physical observables, which are even experimentally accessible, i.e., the physical susceptibilities, which are defined in Eq.~(\ref{equ:sumgeneralizedsusc}). Taking into account only these contributions gives rise to the following approximation for the irreducible vertex, which has been discussed in Ref.~\onlinecite{Tagliavini2018} (see also the supplemental material of Ref.~\onlinecite{Gunnarsson2015} as well as Refs.~\onlinecite{Krien2019} and \onlinecite{Stepanov2019}):



\begin{align}
\label{asymp:Gamma}
\Gamma^{\nu\nu^\prime\omega}_{\text{sp}} \simeq - U  &-\frac{U^2}{2}\chi^{\text{loc}}_{\text{sp}}(\nu^\prime-\nu)  +\frac{U^2}{2}\chi^{\text{loc}}_{\text{ch}}(\nu^\prime-\nu) \nonumber \\ &+ U^2\chi^{\text{loc}}_{\text{pp}}(\omega + \nu+\nu^\prime), 
\end{align}
where $\chi^{\text{loc}}_{\text{sp}}(\omega)$,  $\chi^{\text{loc}}_{\text{ch}}(\omega)$, and $\chi^{\text{loc}}_{\text{pp}}(\omega)$ denote the local spin, charge, and particle-particle (up-down) susceptibility of DMFT, which can be obtained directly from the related AIM. Let us point out that no susceptibility, which depends only on the bosonic Matsubara frequency $\omega$ appears in Eq.(\ref{asymp:Gamma}) since such contribution corresponds to reducible vertex diagrams, which are present in the full\cite{Gunnarsson2015,Tagliavini2018} but not in the irreducible vertex. We also want to remark that the concrete form of Eq.~(\ref{asymp:Gamma}) holds for frequency and time independent interactions in the SU(2) symmetric case.

The approximate form of $\Gamma_{\text{sp}}^{\nu\nu'\omega}$ in Eq.~(\ref{asymp:Gamma}) provides a starting point to investigate the impact of the different local fluctuations of DMFT in the charge, spin, and particle-particle channels on the antiferromagnetic susceptibility and, consequently, on the transition temperature $T_N$ to the antiferromagnetically ordered state. This will be achieved by ``switching'' on and off the various contributions to the local irreducible vertex. To this end we define
\begin{equation}\label{eq:combs}
\Gamma_{\text{sp},w}^{\nu\nu^\prime(\omega=0)} = -U  -w_\text{sp}\frac{1}{2}W_\text{sp}^{\nu^\prime - \nu } +w_\text{ch}\frac{1}{2}W_\text{ch}^{\nu^\prime - \nu }+w_{\text{pp}}\,W_\text{pp}^{\nu^\prime +\nu },
\end{equation}
where for convenience we have introduced the effective screened interactions $W_r^\omega\!=\!U^2\chi_r^\text{loc}(\omega)$ and the binary weights  $w_r =\{0,1\}$ that switch on or off the contributions of the different local fluctuations. Note that we have restricted Eq.~(\ref{eq:combs}) already to the relevant case $\omega\!=\!0$. The above procedure gives rise to $8$ different approximations for $\Gamma_\text{sp}^{\nu\nu'(\omega=0)}$ where this correlation function includes the local fluctuations of DMFT in $0$, $1$, $2$, or all $3$ fluctuation channels. The corresponding numerical results for the $8$ $T_N$ curves for all possible combinations of the three binary weights $w_r$ are discussed in the first part of Sec.~\ref{sec:numerics}.

\subsection{Inverse fluctuations analysis}
\label{sec:invdiag}

An approximation, which is similar to Eq.~(\ref{asymp:Gamma}), has been put forward in Ref.~\onlinecite{Tahvildar-Zadeh1997} where only the second-order contribution (bubble terms) to the local susceptibilities has been taken into account. While we will start from the perturbative expressions of $\chi_r^\text{loc}(\omega)$ we will also consider the fully dressed version of these correlation functions, which contain all local vertex corrections. 

Nevertheless, it is far from obvious what can be expected from such approximations for $\Gamma_{\text{sp}}^{\nu\nu'(\omega=0)}$ beyond the weak-to-intermediate coupling regime. In fact, in this region of the phase diagram the contributions from the fermion-boson triangular vertices\cite{Harkov2021} or the fully irreducible vertices might become non-negligible. To improve our understanding of the AF phase transition also for stronger values of $U$, we propose, as a second step, an inverse or complementary fluctuation analysis of the AF susceptibility where we start from the exact irreducible vertex $\Gamma_{\text{sp}}^{\nu\nu'(\omega=0)}$ of DMFT and {\em subtract} the contributions originating from the local susceptibilities in the charge, spin, and particle-particle channels (note that the spin contribution is negative)
\begin{align}
\label{equ:compdiag}
\widetilde{\Gamma}_{\text{sp},w}^{\nu\nu'(\omega=0)}=\Gamma_{\text{sp}}^{\nu\nu'(\omega=0)}&+w_\text{sp}\frac{1}{2}W_\text{sp}^{\nu^\prime - \nu }-w_\text{ch}\frac{1}{2}W_\text{ch}^{\nu^\prime - \nu }\nonumber\\&-w_{\text{pp}}\,W_\text{pp}^{\nu^\prime +\nu }.
\end{align}
In this way, we keep all nonperturbative low-frequency contributions of the irreducible vertex but we are still able to ``switch'' on and off the high-frequency local fluctuations contained in these correlation functions. The results for this complementary way of the fluctuation analysis will be presented in the second part of Sec.~\ref{sec:numerics}.

\section{Numerical Results}
\label{sec:numerics}

\begin{figure}
    \centering
    \includegraphics[width=0.5\textwidth]{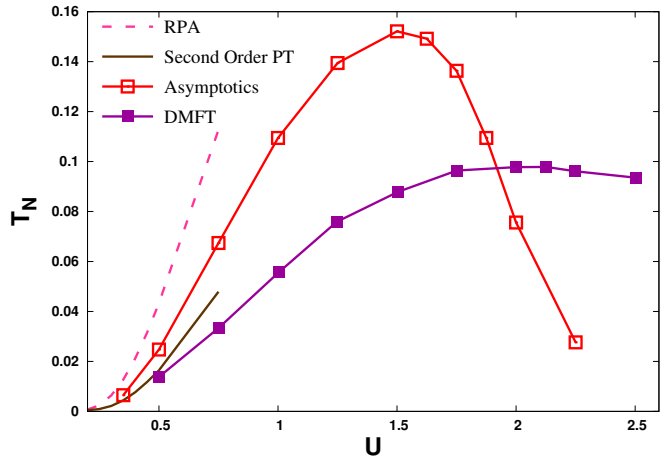}
    \caption{$T_N$ as a function of $U$ calculated with various approaches which differ in the choice of the approximation for $\Sigma(\nu)$ and $\Gamma_\text{sp}^{\nu\nu'\omega}$ (see text). The red empty squares (``Asymptotics'') correspond to Eq.~(\ref{eq:combs}) with all weights $w_r\!=\!1$.}
    \label{fig:weakcoupling}
\end{figure}

In the following, we present the numerical results for the transition temperature $T_N$ to the antiferromagnetically ordered state obtained by the approximations for the local irreducible vertex $\Gamma_\text{sp}^{\nu\nu'\omega}$ of DMFT outlined in the previous section. The DMFT calculations for $\Sigma(\nu)$ and $\Gamma_\text{sp}^{\nu\nu'(\omega=0)}$ have been carried out with exact diagonalization of the related AIM using $4$ bathsites and 160 positive and negative (i.e., in total 320) fermionic Matsubara frequencies for both $\nu$ and $\nu'$. To validate our ED results, we have performed comparisons to corresponding QMC calculations for selected interaction strengths and temperatures, which are presented in Appendix~\ref{app:compQMC}.

The section is organized in the following way: First, in Sec.~\ref{sec:fluctdiag_weakcoupling} we start our analysis for low values of $U$, which allows to compare our results to simple perturbation theory and random phase approximation (RPA) calculations. Second, in Sec.~\ref{sec:fluctdiag_intermediatecoupling}, we will extend our investigations to the entire phase diagram before we finish our analysis with a discussion of the strong-coupling limit in Sec.~\ref{sec:fluctdiag_strongcoupling}.

\subsection{Perturbative analysis}
\label{sec:fluctdiag_weakcoupling}

\begin{figure}
    \centering
    \includegraphics[width=0.5\textwidth]{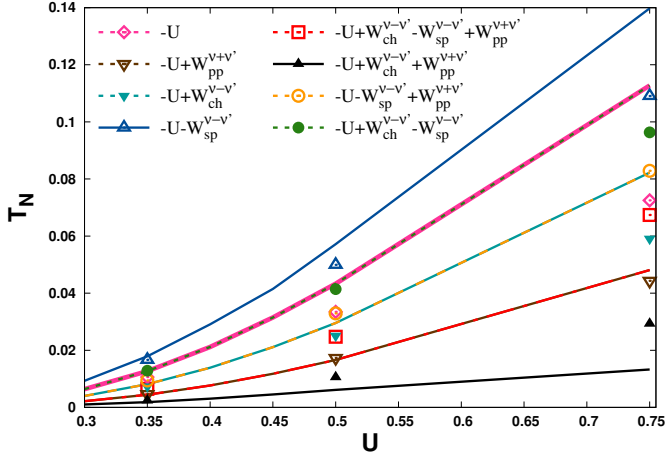}
    \caption{Fluctuation analysis of $T_N$ as a function of $U$ at weak coupling: The different curves correspond to selecting different contributions for $\Gamma_\text{sp}^{\nu\nu'(\omega=0)}$ in Eq.~(\ref{eq:combs}). Lines indicate the result for $\Sigma(\nu)=0$ and $W_r^\omega$ from second-order perturbation theory while the symbols represent calculations including $\Sigma(\nu)$ and the fully dressed $W_r^\omega$ of DMFT containing all local vertex corrections. The pink thick line corresponds to the RPA $T_N$.}
    \label{fig:weakcouplingdiag}
\end{figure}

One of the simplest approaches to calculate $\chi_\text{sp}(\omega,\mathbf{q})$ and $T_N$ for $\frac{U}{t}\ll 1$ is the RPA, a static mean field theory. Within the formal framework discussed in the previous sections, this corresponds to $\Sigma(\nu)\!=\!0$ and $\Gamma_\text{sp}^{\nu\nu'\omega}\!=\!-U$. In three dimensions, RPA gives rise to a $T_N$, which increases exponentially with $U$ as $T_N^\text{RPA}\!\propto\!e^{-\frac{1}{U D(0)}}$ (see dashed pink curve in Fig.~\ref{fig:weakcoupling}) where $D(0)$ is the noninteracting density of states at the Fermi level. A first attempt to understand how local DMFT correlations modify this picture was made in Ref.~\onlinecite{Tahvildar-Zadeh1997} (see also Refs.~\onlinecite{Schauerte2002} and \onlinecite{Keller2001}): There the authors approximated $\Gamma_\text{sp}^{\nu\nu'\omega}$ by a simple second-order diagram, which gave rise to a reduction of $T_N$ by roughly a factor of $3$ (solid brown curve) matching rather well the exact DMFT value (filled violet squares) at small $U$. 
Hence, the authors concluded that the introduction of local DMFT vertex corrections in $\Gamma_\text{sp}^{\nu\nu'\omega}$ leads to a suppression of $T_N$. 

Let us now view this second-order approximation from a more general perspective in the framework of Eqs.~(\ref{asymp:Gamma}) and (\ref{eq:combs}): At second order in $U$, the local spin and charge susceptibilities are equivalent [$\chi_{\text{sp}}^{(2),\text{loc}}(\omega)\!=\!\chi_{\text{ch}}^{(2),\text{loc}}(\omega)\!=\!-\frac{1}{\beta}\sum_\nu G_\text{loc}(\nu)G_\text{loc}(\nu\!+\!\omega)$] and, hence, cancel in the approximate expression for $\Gamma_\text{sp}^{\nu\nu'\omega}$. Thus, only the particle-particle contribution remains, which---due to the above mentioned cancellation---corresponds to either $w_\text{ch}\!=\!w_\text{sp}\!=\!0$ or $w_\text{ch}\!=\!w_\text{sp}\!=\!1$ and $w_\text{pp}\!=\!1$ in Eq.~(\ref{eq:combs}).

We now consider other combinations of the binary weights $w_r$ to gain more insight into the impact of the different local fluctuations on $T_N$, where we again take into account only the second-order contributions for $\chi_r^\text{loc}(\omega)$. In Fig.~\ref{fig:weakcouplingdiag} we can clearly see, that for all combinations of $w_r$ where $w_\text{sp}\!=\!0$ and at least one of the two other weights $w_\text{ch}$ and/or $w_\text{pp}$ equals  $1$ (dashed brown and cyan and solid black lines)  the corresponding $T_N$ is lower than the RPA value (thick pink line). Hence, local charge and particle-particle fluctuations lead to a screening of the bare interaction $U$ in $\Gamma_\text{sp}^{\nu\nu'\omega}$. Interestingly, the reduction of $T_N$ with respect to RPA due to local particle-particle fluctuations (dashed brown line) is stronger than due to charge fluctuations (dashed cyan line) although both  types of fluctuations are equivalent at half filling, i.e., $\chi_\text{ch}^\text{loc}(\omega)\!\equiv\!\chi_\text{pp}^\text{loc}(\omega)$. This can be easily understood from Eqs.~(\ref{asymp:Gamma}) and (\ref{eq:combs}) where the local particle-particle susceptibility enters the vertex $\Gamma_\text{sp}^{\nu\nu'\omega}$ with a factor of $2$ compared to the charge susceptibility.

On the contrary, the combination $w_\text{sp}\!=\!1$ and $w_\text{ch}\!=\!w_\text{pp}\!=\!0$ (solid blue line) leads to a larger $T_N$ with respect to RPA. Hence, local spin fluctuations lead to a ``negative'' screening of $U$ in $\Gamma_\text{sp}^{\nu\nu'\omega}$, which is consistent with the negative sign of $\chi_\text{sp}^\text{loc}(\omega)$ in Eqs.~(\ref{asymp:Gamma}) and (\ref{eq:combs}). Let us stress that no further curves for $T_N$ can be obtained from the three missing combinations of $w_r$ within second-order perturbation theory due to the equivalence of {\em all} three local susceptibilities in this approximation\footnote{Due to the equivalence of all three local susceptibilities the combination $w_\text{sp}\!=\!w_\text{ch}\!=\!w_\text{pp}\!=\!1$ is equivalent to $w_\text{sp}\!=\!w_\text{ch}\!=\!0$, $w_\text{pp}\!=\!1$, the combination $w_\text{sp}\!=\!w_\text{ch}\!=\!1$, $w_\text{pp}\!=\!0$ is equivalent to $w_\text{sp}\!=\!w_\text{ch}\!=\!=w_\text{pp}\!=\!0$ and the combination $w_\text{sp}\!=\!w_\text{pp}\!=\!0$, $w_\text{ch}\!=\!1$ is equivalent to $w_\text{sp}\!=\!w_\text{pp}\!=\!1$, $w_\text{ch}\!=\!0$.}. 

\paragraph*{Beyond perturbation theory.}
In the next step, we go beyond second-order perturbation theory and consider for $\chi_r^\text{loc}(\omega)$ the full local susceptibilities of DMFT including all (local) vertex corrections. Moreover, we take into account the self-energy $\Sigma(\nu)$ to dress the Green's functions in the bare nonlocal susceptibility $\chi_{0,\mathbf{q}}^{\nu\nu'\omega}$ [``bubble'', see Eq.~(\ref{equ:BS})]. The corresponding results for $T_N$ for the various combinations of the binary weights $w_r$ are presented with different symbols in Fig.~\ref{fig:weakcouplingdiag} where for each of the $8$ sets of $w_r$ we have used the same color as in the perturbative treatment. Not surprisingly, at the lowest value of $U\!=\!0.35$, the results of second-order perturbation theory (lines) almost coincide with the the ones where the exact DMFT self-energy and local susceptibilities have been used. Upon increasing $U$, they start to differ lifting also the degeneracies, which occur in the perturbative treatment due to the equivalence of all local susceptibilities within second-order perturbation theory.

\begin{figure}
    \centering
    \includegraphics[width= \columnwidth]{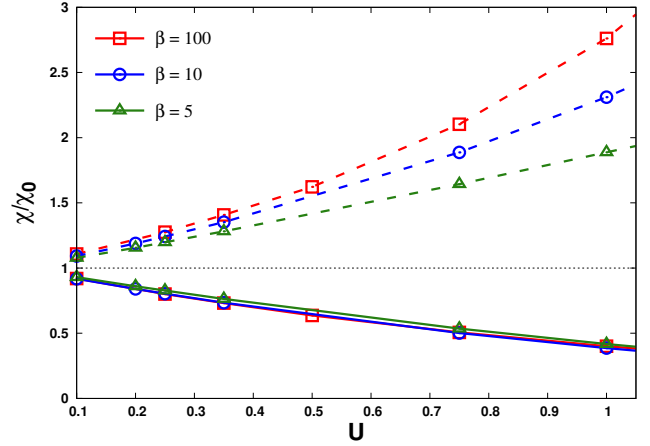}
    \caption{Local density (thick lines) and magnetic (dashed lines) susceptibilities $\chi_\text{ch}^\text{loc}(\omega\!=\!0)$ and $\chi_\text{sp}^{\text{loc}}(\omega\!=\!0)$ renormalized by their noninteracting values as a function of the interaction strength for three different temperatures. }
    \label{fig:chi_loc}
\end{figure}

First, we observe that the introduction of the self-energy for $\Gamma_\text{sp}^{\nu\nu'\omega}\!=\!-U$ (pink diamonds, $w_r\!=\!0$ for all channels) leads to a clear reduction of $T_N$ with respect to the RPA result (thick pink curve). This is the expected behavior since the DMFT self-energy suppresses the spectral weight at the Fermi level and, hence, reduces the bubble term in the BS Eq.~(\ref{equ:BS}). It is obvious that this effect also plays a role for the other combinations of $w_r$ where vertex corrections in $\Gamma_\text{sp}^{\nu\nu'\omega}$ are taken into account. However, the situation is more complicated due to the additional dressing of the local susceptibilities in Eqs.~(\ref{asymp:Gamma}) and (\ref{eq:combs}). In fact, the vertex corrections in $\chi_\text{ch}^\text{loc}(\omega)$ and $\chi_\text{pp}^\text{loc}(\omega)$ lead to a suppression of these fluctuations with respect to second-order perturbation theory while the exact local spin susceptibility is larger than its second-order counterpart (see Fig.~\ref{fig:chi_loc}). This leads to a dominance of local spin fluctuations in $\Gamma_\text{sp}^{\nu\nu'\omega}$ in Eqs.~(\ref{asymp:Gamma}) and (\ref{eq:combs}) and, due to the negative sign with which this contribution enters the vertex, provides a ``negative'' screening, i.e., an enhancement of this vertex with respect to RPA or simple second-order perturbation theory. This leads to a competition of local correlation effects on the one- and the two-particle level: The former, expressed by the self-energy, reduce $\chi_\text{sp}(\omega,\mathbf{q})$ and $T_N$ while the latter try to enhance it with respect to perturbation theory. Which of both effects ``wins'' is reflected in the larger or lower value of $T_N$ within the perturbative (lines) with respect to the DMFT treatment (points) in Fig.~\ref{fig:weakcouplingdiag}. The precise hierarchy of the curves for different $w_r$ then strongly depends on the exact interplay between $\Sigma(\nu)$ and $\chi_r^\text{loc}(\omega)$. 


\subsection{Complete phase diagram}
\label{sec:fluctdiag_intermediatecoupling}
\begin{figure}
\begin{center}
\includegraphics[width = \columnwidth]{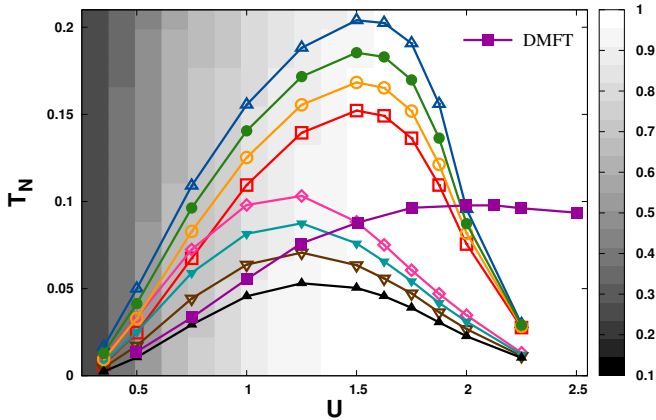}
\caption{ The different symbols represent the critical temperature curves obtained using the 8 possible combinations of the $w_r$ for the vertex function in Eq.(\ref{eq:combs}), the solid lines being a guide to the eye. The color/symbol legend is the same as in Fig.(\ref{fig:weakcouplingdiag}). The exact N\'eel temperature of DMFT is shown as a reference (filled violet squares).  The gray intensity map in the background encodes the ratio  $-\chi_{\uparrow\downarrow}(\omega = 0)/\chi_{\uparrow\uparrow}(\omega = 0)$ (see the main text for more details).}
\label{fig:diagn}
\end{center}
\end{figure}
Figure~\ref{fig:diagn} shows the fluctuation analysis of $T_N$ in a larger $U$ region where the same colors and symbols for the different combinations of the $w_r$'s have been used as in the weak-coupling analysis in Fig.~\ref{fig:weakcouplingdiag}. For $U\!\ge\!1$, we observe a clear hierarchy of the curves for $T_N$ calculated with the different approximations of $\Gamma_\text{sp}^{\nu\nu'\omega}$: All results for the transition temperature obtained by setting $w_\text{sp}\!=\!0$ (filled cyan lower triangles, empty brown lower triangles, filled black upper triangles) lie below the pink curve (empty diamonds), which represents the result for $T_N$ without vertex corrections [$\Gamma_\text{sp}^{\nu\nu'\omega}\!=\!-U$, $w_r\!=\!0$ for all channels $r$ in Eq.~(\ref{eq:combs})]. This is consistent with the weak-coupling result that local charge and particle-particle fluctuations in $\Gamma_\text{sp}^{\nu\nu'\omega}$ lead to a screening of the bare interaction $-U$ in this vertex function and, hence, reduce the transition temperature with respect to to $\Gamma_\text{sp}^{\nu\nu'\omega}\!=\!-U$. This reduction is obviously largest when both (i.e., charge and particle-particle) screening channels contribute to $\Gamma_\text{sp}^{\nu\nu'\omega}$ ($w_\text{ch}\!=\!w_\text{pp}\!=\!1$, filled black upper triangles in Fig.~\ref{fig:diagn}). Although at half filling local charge and particle-particle fluctuations are equivalent, the latter (empty brown lower triangles) suppress $T_N$ stronger than the inclusion of charge fluctuations (filled cyan lower triangles). As already pointed out in the weak-coupling analysis, this can be ascribed to the different prefactors with which $\chi_\text{pp}^\text{loc}$ and $\chi_\text{ch}^\text{loc}$ enter Eqs.~(\ref{asymp:Gamma}) and (\ref{eq:combs}).

Conversely, we find the opposite effect when spin fluctuations are included: For $U\!\ge\!1$, all curves for $T_N$ in Fig.~\ref{fig:diagn} where $w_\text{sp}\!=\!1$ (empty blue upper triangles, filled green circles, empty orange circles, empty red squares) lie above the pink diamonds for which all vertex corrections for $\Gamma_\text{sp}^{\nu\nu'\omega}$ are neglected. Hence, local spin fluctuations provide a ``negative'' screening, which enhances the bare interaction and, consequently, the antiferromagnetic susceptibility and $T_N$ with respect to the calculation with $\Gamma_\text{sp}^{\nu\nu'\omega}\!=\!-U$. The situation where {\em only} local spin fluctuations are considered (i.e., $w_\text{sp}\!=\!1$ and $w_\text{ch}\!=\!w_\text{pp}\!=\!0$, empty blue upper triangles) obviously gives rise to the largest $T_N$ while the inclusion of one or both of the two other channels  leads to a smaller value of $T_N$. However, for $U\!\ge\!1$, the screening provided by the local charge and particle-particle fluctuations is always smaller than the enhancement of $\Gamma_\text{sp}^{\nu\nu'\omega}$ due to local spin fluctuations.

The above considerations are obviously an effect of the strong enhancement of local spin fluctuations with respect to local charge and particle-particle fluctuations upon increasing $U$ within DMFT. Hence, it is interesting to analyze the relative difference between $\chi^\text{loc}_\text{sp}$ and $\chi^\text{loc}_\text{ch}$ (which is equivalent to $\chi^\text{loc}_\text{pp}$ at half filling) as a function of $U$. This is shown in Fig.~\ref{fig:diagn} by means of a gray-scale intensity map that encodes the relative deviation between the static ($\omega\!=\!0$) charge and spin fluctuations defined as $\frac{\chi_s(\omega = 0) - \chi_c(\omega = 0)}{\chi_s(\omega = 0) + \chi_c(\omega = 0)} = -\frac{\chi_{\uparrow\downarrow}(\omega = 0)}{\chi_{\uparrow\uparrow}(\omega = 0)}$. At $U\!=\!0$, this ratio vanishes exactly while for $U\!\rightarrow\!\infty$ it approaches unity. Hence, when the relative deviation increases (brighter gray shades in Fig.~\ref{fig:diagn}), spin fluctuations in $\Gamma_\text{sp}^{\nu\nu'\omega}$ gradually become dominant. Consistently with the discussion above, such a dominance of spin fluctuations is observed already in the weak-to-intermediate coupling regime at $U\!\ge\!1$ where the relative deviation 
is about 80\% (although this is well below the Mott local moment regime, which sets in at $U\!\sim\!2.3$). 

Let us briefly comment on the general structure of $T_N$ as a function of $U$ in Fig.~\ref{fig:diagn}. We observe a qualitatively similar behavior as for the exact DMFT phase transition (filled violet squares): At small values of $U$, $T_N$ increases, then reaches a maximum at intermediate coupling and decreases at large values of $U$. However, on a quantitative level it is quite clear that Eqs.~(\ref{asymp:Gamma}) and (\ref{eq:combs}) do not provide a  good approximation for $\Gamma_\text{sp}^{\nu\nu'\omega}$ to calculate $T_N$. In fact, the maxima of $T_N$ for the approximate calculations are shifted with respect to the DMFT result and, moreover, the vertex in Eq.(\ref{eq:combs}) fails completely to recover the strong-coupling limit where $T_N \propto t^2/U$. In this regime, the bubble goes to zero too rapidly and even in the case with $w_s\!=\!1$ (and $w_\text{ch}^\text{loc}\!=\!w_\text{pp}^{\text{loc}}\!=\!0)$, the formation of local moments  encoded in $\chi_s(\omega)$ it is not sufficient to balance the loss of coherence occurring in the bubble via the insertion of the local self-energy in the Green's function.

This indicates that beyond weak coupling contributions to the irreducible vertex $\Gamma_\text{sp}^{\nu\nu'\omega}$ become important, which are not considered in Eqs.~(\ref{asymp:Gamma}) and (\ref{eq:combs}). Within the dual fermion approach, a good quantitative agreement has been achieved in the framework of the single-boson exchange formalism\cite{Harkov2021}, which includes additional contributions in the expansion, in particular the triangular vertices, which account for the coupling between the electrons and the collective modes. In the latter situation the approximations have been performed in a dual space to the full (instead of the irreducible) local vertex of DMFT. The inclusion of triangular vertices also in the approximation for $\Gamma_\text{sp}^{\nu\nu'\omega}$ in Eq.~(\ref{asymp:Gamma}) would be definitely interesting and can potentially improve our results. However, in this paper we follow the alternative route outlined in Sec.~\ref{sec:invdiag}
leaving the question of the triangular vertices for future research work.

\begin{figure}
    \centering
    \includegraphics[width=0.5\textwidth]{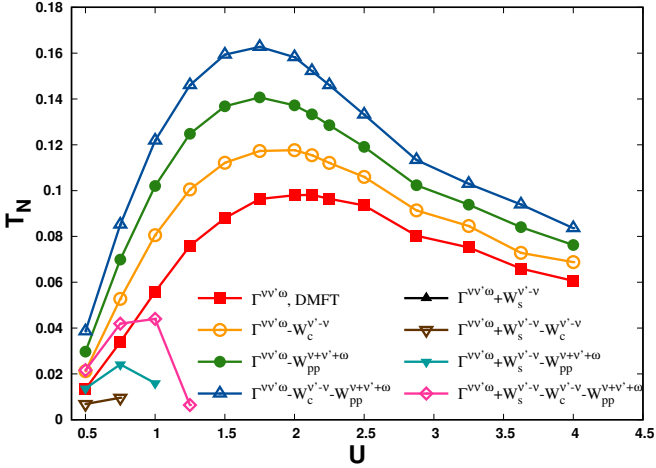}
    \caption{Complementary fluctuations analysis. The different curves for $T_N$ are obtained by subtracting the local charge, spin, and particle-particle susceptibilities from the exact DMFT vertex $\Gamma_\text{sp}^{\nu\nu'\omega}$ [see Eq.~\ref{equ:compdiag}].}
    \label{fig:compfluc}
\end{figure}

\paragraph*{Complementary fluctuation analysis.} To obtain a fluctuation analysis of $T_N$, which is quantitatively more similar to the DMFT transition curve, we proceed with Eq.~(\ref{equ:compdiag}). There, instead of constructing the vertex $\Gamma_\text{sp}^{\nu\nu'\omega}$ just from its asymptotic contributions, we subtract these terms from the exact DMFT expression for $\Gamma_\text{sp}^{\nu\nu'\omega}$ to quantify the effects of the different local DMFT fluctuations on $T_N$. The results are presented in Fig.~\ref{fig:compfluc}.

Consistent with the previous findings, the curves where charge (empty orange circles), particle-particle (filled green circles) or both (empty blue upper triangles) types of fluctuations are subtracted from the DMFT $\Gamma_\text{sp}^{\nu\nu'\omega}$ are enhanced with respect to the exact DMFT result (red filled squares). Hence, these channels indeed provide a screening of the bare interaction $-U$ and removing them leads to an enhancement of $T_N$ with respect to the DMFT result. Obviously, subtracting both the charge and the particle-particle local susceptibilities (empty blue upper triangles) gives rise to the largest transition temperature. On the other hand, the removal of $\chi_\text{pp}^\text{loc}$ (filled green circles) leads to a stronger enhancement than the subtraction of $\chi_\text{ch}^\text{loc}$ (empty orange circles), which, again, is explained by the different prefactors with which these two types of fluctuations enter Eq.~(\ref{equ:compdiag}). Interestingly, there is a clear enhancement of $T_N$ of all three curves with respect to DMFT up to the largest considered couplings although local charge and particle-particle fluctuations are already exponentially suppressed in this parameter regime.

The picture changes completely when we subtract the local spin fluctuations (empty pink diamonds, filled cyan lower triangles, empty brown lower triangles, filled black upper triangles in Fig.~\ref{fig:compfluc}). In this case, $T_N$ is in general strongly suppressed with respect to the exact DMFT result (filled red squares) and even vanishes at a small finite $U$. If we remove {\em only} $\chi_\text{sp}^\text{loc}$ we do not find a finite transition temperature at all in the considered parameter regime (i.e., there is no curve corresponding to filled black upper triangles in Fig.~\ref{fig:compfluc}). By considering the additional removal of one or both of the two other channels (charge and particle-particle), a finite transition temperature emerges at small values of $U$ but vanishes upon increasing the interaction strength up to $U\!\sim\!1.25$. Note that this value  is still located within the metallic region well below the Mott transition and the local moment regime at $U\!\sim\!2.3$. Nevertheless, local spin fluctuations in $\Gamma_\text{sp}^{\nu\nu'\omega}$ are already an indispensable ingredient to obtain a transition to the antiferromagnetically ordered state. 

\subsection{Strong-coupling limit}
\label{sec:fluctdiag_strongcoupling}

\par 

Figure~\ref{fig:compfluc} suggests that the three curves, where local charge and/or particle-particle fluctuations are subtracted (blue upper empty triangles, green filled circles, orange empty circles), approach the DMFT transition line in the limit $U\!\rightarrow\!\infty$. This is reasonable since local charge and particle-particle fluctuations are exponentially suppressed in the large coupling limit. To verify this intuitive argument, we consider the following useful approximation for the calculation of $T_N$ at large $U$, which recovers the expected power-law behavior of the critical temperature.

\begin{figure}
    \centering
    \includegraphics[width =0.9 \columnwidth]{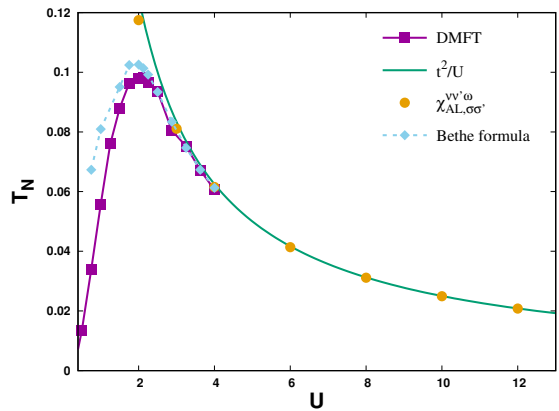}
    \caption{Critical temperature as a function of the interaction in the strong-coupling approximation (filled orange circles) compared with the DMFT data (filled violet squares). Blue diamonds indicate $T_N$ as obtained from Eq.~(\ref{eq:bse:bethe}).}
    \label{fig:strong_coupl}
\end{figure}

Solving Eq.~(\ref{equ:BS}) for the (inverse of the) generalized spin susceptibility $\chi_\mathbf{\text{sp},q}^{\nu\nu'\omega}$ yields
\begin{equation}\label{eq:bse_rewra}
   \bar{\bar{\chi}}^{-1}_\text{sp}(\omega,\bq)=  \bar{\bar{\chi}}^{-1}_0(\omega,\bq) + \frac{1}{\beta^2}\Gamma_\text{sp}^{\nu\nu'\omega}, 
\end{equation}
where we chose a matrix notation in the fermionic Matsubara frequencies, i.e., $\bar{\bar{\chi}}(\omega,\bq)\!=\!\left[\bar{\bar{\chi}}(\omega,\bq)\right]_{\nu\nu^\prime}\!\equiv\!\chi^{\nu\nu^\prime \omega}_\bq$ and ``$^{-1}$'' denotes the inversion of this (infinite) matrix. The local irreducible vertex $\Gamma_\text{sp}^{\nu\nu'\omega}$, in turn, can be obtained from a purely local version of the BS Eq.~(\ref{equ:BS}), which reads
\begin{equation}\label{eq:bse_rewrb}
   \frac{1}{\beta^2}\Gamma_\text{sp}^{\nu\nu'\omega}=\bar{\bar{\chi}}^{-1}_\text{sp}(\omega)-\bar{\bar{\chi}}^{-1}_0(\omega) , 
\end{equation}
\affiliation{}where $\bar{\bar{\chi}}^{-1}_\text{sp}(\omega)$ is the local generalized spin susceptibility of the auxiliary AIM and $\bar{\bar{\chi}}_0(\omega)=\chi_0^{\nu\nu'\omega}\!=\!-\beta G(\nu)G(\nu+\omega)\delta_{\nu\nu'}$ is the local bubble. Combining Eqs.~(\ref{eq:bse_rewra}) and (\ref{eq:bse_rewrb}) then yields
\begin{equation}\label{eq:bse_rewr}
   \bar{\bar{\chi}}^{-1}_\text{sp}(\omega,\bq)=  \bar{\bar{\chi}}^{-1}_0(\omega,\bq) - \bar{\bar{\chi}}_{0}^{-1}(\omega) + \bar{\bar{\chi}}^{-1}_{\text{sp}}(\omega).
\end{equation}
The AF susceptibility corresponds to $\omega\!=\!0$ and $\mathbf{q}\!=\!\boldsymbol{\Pi}$ in this equation. Due to the DMFT self-consistency condition, the nonlocal bubble term reduces to $\bar{\bar{\chi}}_0(\omega\!=\!0,\mathbf{q}\!=\!\boldsymbol{\Pi}) = -\beta \mathbbm{1}\, D(\zeta_\nu)/\zeta_\nu$, where we have introduced the Hilbert transform $D(\zeta) = \int_{-\infty}^{+\infty}d\epsilon g(\epsilon)[\zeta-\epsilon]^{-1} $ of the noninteracting density of states $g(\epsilon)$, with $\zeta_\nu = i \nu + \mu - \Sigma(\nu)$ ($\mathbbm{1}\!=\!\delta_{\nu\nu'}$ is the unit matrix in the $\nu$-$\nu'$ space). The local bubble, on the other hand, can be expressed in terms of the Hilbert transform as $\bar{\bar{\chi}}_0(\omega = 0) = -\beta \mathbbm{1}\,D^2(\zeta_\nu)$. In the specific case of a semi-circular density of states $g(\epsilon) = \frac{1}{2 \pi t^2}\sqrt{\epsilon^2-4t^2}$ (Bethe lattice), the contribution arising from the bubble terms simplifies considerably and Eq.(\ref{eq:bse_rewr}) reduces to
\begin{equation}\label{eq:bse:bethe}
  \bar{\bar{\chi}}_\text{sp}^{-1}(0,\boldsymbol{\Pi})=   \bar{\bar{\chi}}^{-1}_{\text{sp}}(0) - \frac{t^2}{\beta}\,\mathbbm{1}.
\end{equation}
We recall that a very similar formula is found for the
homogeneous case {\bfseries q} = 0 \cite{Georges1996,delre2018,Reitner2020} where
where $-t^2$ is replaced with $+t^2$ . To make contact to the $3d$ lattice model considered before, we choose $t$ in such a way that twice the second moment of the density of states $g(\epsilon)$ of the Bethe lattice is normalized to $D\!=\!1$ (as for the $3d$ cubic lattice). Here, this corresponds to $D\!=\!2t\!=\!1$, i.e., $t\!=\!\sqrt{6}t_\text{3d}$. With this choice both lattice types will lead to quantitatively same results for the local correlation functions of DMFT since the DMFT self-consistency is mainly controlled by the second moment of the non-interacting density of states\cite{Bulla2000}. For the nonlocal spin susceptibility and the related $T_N$ this also holds at  strong coupling where these quantities are mainly governed by the local DMFT self-energy and irreducible vertex functions. This can be seen in Fig.~\ref{fig:strong_coupl} where the dashed blue line, which corresponds to $T_N$ as obtained from the expression in Eq.~(\ref{eq:bse:bethe}) for the Bethe lattice (but with the local correlation functions obtained from a DMFT calculation for the simple cubic lattice), agrees very well with the exact DMFT calculation for the simple cubic lattice (filled violet squares). On the other hand, at weaker coupling $T_N$ is mainly determined by the noninteracting density of states ($T_N^\text{RPA}\!\propto\!e^{-\frac{1}{U D(0)}}$) at zero energy, which is in general different for the Bethe and the simple cubic lattice when their second moments are fixed to the same value. This can explain the different results for $T_N$ from DMFT and the Bethe lattice treatment in this parameter regime.


\par Starting from Eq.(\ref{eq:bse:bethe}), an obvious approximation for the strong coupling limit consists in replacing $\bar{\bar{\chi}}_{\text{sp}}(0)$ by the corresponding generalized susceptibility calculated in the atomic limit $\bar{\bar{\chi}}^{\text{AL}}_\text{sp}(0)$. The inverse of $\bar{\bar{\chi}}_{\text{sp}}^{\text{AL}}(0)$, which is required in 
Eq.~(\ref{eq:bse:bethe}), has been calculated analytically (at half filling) in Ref.~\onlinecite{Thunstr2018}. The same methods, which have been used for the inversion of the $\bar{\bar{\chi}}_\text{sp}^{\text{AL}}(\omega)$ in this paper, can be applied to invert $ \bar{\bar{\chi}}_\text{sp}^{-1}(0,\boldsymbol{\Pi})$ in Eq.~(\ref{eq:bse:bethe}). The transition temperature $T_N$ is then determined by the condition that the matrix $\bar{\bar{\chi}}_{\text{sp}}^{-1}(0,\boldsymbol{\Pi})$ is singular, corresponding to a divergence of  $\bar{\bar{\chi}}_{\text{sp}}(0,\boldsymbol{\Pi})$. From this condition we obtain the following expression of $T_N$ in terms of  $T$, $U$, and $t$ (for details see Appendix~\ref{app:alanalytic}):
\begin{equation}
\label{equ:alanalytic}
T_N=\frac{t^2}{U}\frac{1}{1+\frac{t^2}{U^2}}+O\left(e^{-\sqrt{1-\frac{4t^2}{U^2}}\frac{\beta U}{2}}\right)
\end{equation}
Considering only the leading order in $t/U$, this equation obviously reproduces the large $U$ behavior of $T_N$ given by $T_N\sim t^2/U$.
This is also illustrated in Fig.~\ref{fig:strong_coupl}, where we show that the N\'eel temperature, calculated within the AL approximation (filled orange circles), coincides with the exact DMFT result (filled violet squares) for $U\gtrsim2.5$ and reproduces the correct large coupling limit $T_N\sim t^2/U$ (green line).

To perform the fluctuation analysis of $T_N$ for $U\rightarrow\infty$, we consider the susceptibilities in the atomic limit
\begin{subequations}
\label{equ:suscal}
\begin{align}
&\chi_\text{ch}^\text{loc}(\omega)=\chi_\text{pp}^\text{loc}(\omega)=\frac{\beta}{2}\frac{1}{1+e^{\frac{\beta U}{2}}}\delta_{\omega 0}\\
&\chi_\text{sp}^\text{loc}(\omega)=\frac{\beta}{2}\frac{1}{1+e^{-\frac{\beta U}{2}}}\delta_{\omega 0}.
\end{align}
\end{subequations}
We can see that $\chi_\text{ch}^\text{loc}$ and $\chi_\text{pp}^{\text{loc}}$ vanish exponentially with $U$. Since we have neglected such terms in Eq.~(\ref{equ:alanalytic}), there is no contribution from these fluctuations to $T_N$ in the large coupling limit and, hence, all lines of the complementary fluctuation analysis in Fig.~\ref{fig:compfluc}, where $\chi_\text{sp}^\text{loc}$ has not been subtracted, should collapse on the DMFT curve for $U\rightarrow\infty$. It is, however, interesting that even for the largest $U$ value, where the exponential suppression of local charge and particle-particle susceptibilities is already rather strong, the difference between the curves is still sizable. Indeed, if we would have used the corresponding local charge and particle-particle susceptibilities of the AL (instead of the exact DMFT ones) the curves would have already collapsed onto the DMFT line at $U~\sim 3.0$. This indicates that local DMFT charge and particle-particle fluctuations cannot completely be neglected in this parameter regime. To quantify this statement we have calculated the ratio $\chi_\text{ch}^\text{loc}(\omega\!=\!0)/\chi_\text{sp}^\text{loc}(\omega\!=\!0)$ in both the DMFT and the AL. At the largest value of the interaction ($U\!=\!4$) and at a temperature  close to the phase transition ($\beta\!=\!16$), this ratio is given by $5.5\times 10^{-4}$ within DMFT. This is indeed one order of magnitude smaller than the absolute difference between $T_N$ of DMFT (red filled squares in Fig.~\ref{fig:compfluc}) and the transition temperature obtained by removing charge fluctuations (orange empty circles in Fig.~\ref{fig:compfluc}). The corresponding ratio calculated using the same set of parameters but in the AL from Eqs.~(\ref{equ:suscal}) yields a value of $1.3\times 10^{-14}$, which is about 10 orders of magnitudes smaller than the DMFT result.

\par It is worth to notice that it is highly important {\em how} the AL approximation is performed in our calculations. In fact, we did not replace the irreducible vertex $\Gamma_\text{sp}^{\nu\nu'\omega}$ with the one of the atomic limit, which would not reproduce the correct behavior of $T_N$ at strong coupling. This is similar to the two options how $T_N$ of DMFT for large $U$ can be calculated within the dual fermion formalism as it has been discussed in Ref.~\onlinecite{Krien2019a}. There, two types of approximations have been performed for the full (instead of the irreducible) vertex. Again only one of these approximations produces the correct large $U$ behavior of $T_N$. How this is linked to the approximations presented in this paper is an interesting open question for future research work.


\section{Momentum dependence of the spin susceptibility}
\label{sec:spinsusc}

\begin{figure}
    \centering
    \includegraphics[width=\columnwidth]{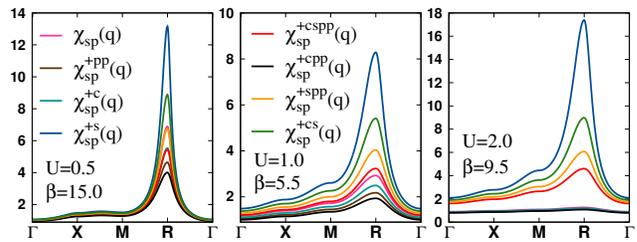}
    \caption{Static lattice spin susceptibility $\chi_{\text{sp}}(\omega\!=\!0,\mathbf{q})$ as obtained from the direct fluctuation analysis (see Secs.~\ref{sec:directdiag} and \ref{sec:fluctdiag_weakcoupling}), where local charge(+c), spin(+s), and particle-particle(+pp) fluctuations are added to $\Gamma_{\text{sp}}^{\nu\nu'\omega}=-U$ in the BS Eq.~(\ref{equ:BS}), for three different interaction values and temperatures along a high symmetry path [$\Gamma$-X-M-R-$\Gamma$=$(0,0,0)$-$(\pi,0,0)$-$(\pi,\pi,0)$-$(\pi,\pi,\pi)$-$(0,0,0)$] in the Brillouin zone of the simple cubic lattice. The color code for the curves, corresponding to the consideration of different combinations of the local correlation functions $\chi_\text{ch}^\text{loc}$, $\chi_\text{sp}^\text{loc}$, and $\chi_\text{pp}^\text{loc}$ for $\Gamma_{\text{sp}}^{\nu\nu'\omega}$ [see Eqs.~(\ref{asymp:Gamma}) and (\ref{eq:combs})], is the same as in Figs.~\ref{fig:weakcouplingdiag} and \ref{fig:diagn}.} 
    \label{fig:spinsuscq_direct}
\end{figure}

The N\'eel temperature $T_N$ is obtained from the static (i.e., $\omega\!=\!0)$ lattice spin susceptibility $\chi_{\text{sp}}(\omega,\mathbf{q})$ at $\mathbf{q}\!=\!\mathbf{\Pi}$ as defined in Eq.~(\ref{equ:defafspinsusc}). To gain further insights into the different approximations for this response function, which originate from the presence or absence of local charge, spin, and particle-particle fluctuations in the irreducible vertex $\Gamma_{\text{sp}}^{\nu\nu'\omega}$, we have also analyzed the momentum dependence of $\chi_{\text{sp}}(\omega\!=\!0,\mathbf{q})$. 

The results for the fluctuation analysis in the weak-coupling regime (see Sec.~\ref{sec:fluctdiag_weakcoupling}) are displayed in Fig.\ref{fig:spinsuscq_direct},  where  $\chi_{\text{sp}}(\omega\!=\!0,\mathbf{q})$ is shown as a function of the momentum $\mathbf{q}$ along a high-symmetry path in the Brillouin zone of the simple cubic lattice for three interaction values $U$. The different curves correspond to the inclusion of different combinations of the local correlation functions $\chi_\text{ch}^\text{loc}$, $\chi_\text{sp}^\text{loc}$, and $\chi_\text{pp}^\text{loc}$  into $\Gamma_\text{sp}^{\nu\nu'\omega}$ [see Eq.~(\ref{eq:combs})]. For each $U$ value we have selected a temperature above the highest $T_N$ in Fig.~\ref{fig:diagn} (empty blue upper triangles), which is obtained by including  solely $\chi_\text{sp}^\text{loc}$  into $\Gamma_\text{sp}^{\nu\nu'\omega}$ [i.e., $w_\text{ch}=w_\text{pp}=0$ and $w_\text{sp}=1$ in Eq.~(\ref{eq:combs})]. As expected, we observe the same order of the curves as in Figs.~\ref{fig:weakcouplingdiag} and \ref{fig:diagn}. The lattice spin susceptibility $\chi_\text{sp}^{+\text{s}}(\omega\!=\!0,\mathbf{q})$, where only local spin fluctuations have been included in $\Gamma_\text{sp}^{\nu\nu'\omega}$ [blue lines, $w_\text{ch}\!=\!w_\text{pp}\!=\!0$ and $w_\text{sp}\!=\!$ in Eq.~(\ref{eq:combs})] is larger than $\chi_\text{sp}(\omega\!=\!0,\mathbf{q})$, where $\Gamma_\text{sp}^{\nu\nu'\omega}\!=\!-U$ (pink line) for all values of $U$. This demonstrates that local spin fluctuations always lead to an enhancement the corresponding nonlocal spin susceptibility. On the contrary, including local charge (cyan lines), particle-particle (brown lines), or both types (black lines) of fluctuations in $\Gamma_\text{sp}^{\nu\nu'\omega}$ leads to a decrease of the corresponding lattice spin susceptibility with respect to $\Gamma_\text{sp}^{\nu\nu'\omega}\!=\!-U$ (pink line). If we include a combination of local spin and local charge and/or particle-particle fluctuations in $\Gamma_\text{sp}^{\nu\nu'\omega}$ (green, brown, and red lines) the situation is different for weak and intermediate-to-strong coupling: For $U\!=\!0.5$ (left panel in Fig.~\ref{fig:spinsuscq_direct}), the local fluctuations in all three scattering channels are of the same order of magnitude and, hence, the order of the lines depends on the detailed interplay between these three local correlation functions.

On the contrary, for $U\!=\!1.0$ and $U\!=\!2.0$ the local spin fluctuations are considerably larger than the corresponding local charge and particle-particle fluctuations which gives rise to a definite hierarchy of curves in the middle and right panels of Fig.~\ref{fig:spinsuscq_direct}: The lattice spin susceptibilities $\chi_\text{sp}^{+r}(\omega=0,\mathbf{q})$, for which the local spin fluctuation are contained in $\Gamma_\text{sp}^{\nu\nu'\omega}$ (blue, green, orange and red lines), are always larger than the corresponding lattice spin susceptibilities with $\Gamma_\text{sp}^{\nu\nu'\omega}\!=\!-U$ Consistently with the results of the previous sections, the largest value is obtained when solely the spin fluctuations are included in the $\Gamma_\text{sp}^{\nu\nu'\omega}$ (blue curve), while the additional consideration of local charge (green lines), particle-particle (orange lines), or both types (red lines) of fluctuations leads to a suppression of the lattice spin susceptibility with respect to the (blue) spin-only curve. 


Interestingly, the difference between the curves is largest at the R point ($\mathbf{q}=\mathbf{\Pi}$), decreases with increasing distance from this antiferromagnetic wave vector and becomes comparatively small at the $\Gamma$ point [$\mathbf{q}\!=\!(0,0,0)$]. This means that the suppression of $\chi_\text{sp}(\omega\!=\!0,\mathbf{q})$ due to local charge and particle-particle fluctuations and the enhancement due to local spin fluctuations is strongest at $\mathbf{q}\!=\!\mathbf{\Pi}$ while it becomes rather moderate away from this point. This observation is quite remarkable since it implies that a {\em purely local} modification of the {\em local} DMFT vertex $\Gamma_\text{sp}^{\nu\nu'\omega}$ gives rise to a {\em momentum dependent} modification of $\chi_\text{sp}(\omega\!=\!0,\mathbf{q})$.

\begin{figure}
    \centering
    \includegraphics[width=\columnwidth]{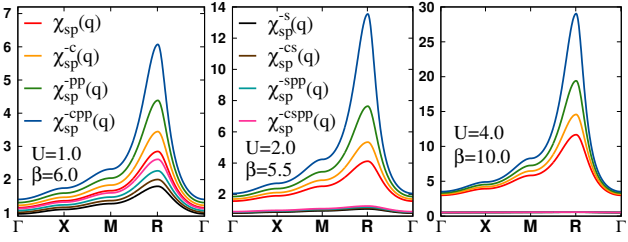}
    \caption{Static lattice spin susceptibility $\chi_{\text{sp}}(\omega\!=\!0,\mathbf{q})$ as obtained from the inverse fluctuation analysis (see Secs.~\ref{sec:invdiag} and \ref{sec:fluctdiag_intermediatecoupling}), where local charge(-c), spin(-s), and particle-particle(-pp) fluctuations are subtracted from $\Gamma_{\text{sp}}^{\nu\nu'\omega}$ in the BS Eq.~(\ref{equ:BS}), for three different interaction values and temperatures along a high-symmetry path [$\Gamma$-X-M-R-$\Gamma$=$(0,0,0)$-$(\pi,0,0)$-$(\pi,\pi,0)$-$(\pi,\pi,\pi)$-$(0,0,0)$] in the Brillouin zone of the simple cubic lattice. The color code for the curves, corresponding to the subtraction of different combinations of the local correlation functions $\chi_\text{ch}^\text{loc}$, $\chi_\text{sp}^\text{loc}$, and $\chi_\text{pp}^\text{loc}$ from $\Gamma_{\text{sp}}^{\nu\nu'\omega}$, is the same as in Fig.~\ref{fig:compfluc}.} 
    \label{fig:spinsuscq_comp}
\end{figure}

We have also calculated the momentum dependence of the spin susceptibility corresponding to the inverse fluctuation analysis where local fluctuations are gradually subtracted from the exact $\Gamma_\text{sp}^{\nu\nu'\omega}$ of DMFT (cf. Secs.~\ref{sec:invdiag} and \ref{sec:fluctdiag_intermediatecoupling} and Fig.~\ref{fig:compfluc}). 
The results are presented in Fig.~\ref{fig:spinsuscq_comp} where the different curves correspond to the {\em removal} of different combinations of the local correlation functions $\chi_\text{ch}^\text{loc}$, $\chi_\text{sp}^\text{loc}$, and $\chi_\text{pp}^\text{loc}$ from the exact $\Gamma_\text{sp}^{\nu\nu'\omega}$ of DMFT [see Eq.~(\ref{equ:compdiag})]. For each value of $U$ we have selected a temperature above the highest $T_N$ in Fig.~\ref{fig:compfluc} (empty blue upper triangles), which is obtained by subtracting $\chi_\text{ch}^\text{loc}$ and $\chi_\text{pp}^\text{loc}$ from $\Gamma_\text{sp}^{\nu\nu'\omega}$ [i.e., $w_\text{ch}=w_\text{pp}=1$ and $w_\text{sp}=0$ in Eq.~(\ref{equ:compdiag})].
Analogously to the weak coupling analysis, the hierarchy of the different curves corresponds to the one in Fig.~\ref{fig:compfluc} for all three values of $U$: The lattice spin susceptibilities $\chi_\text{sp}^{-r}(\omega=0,\mathbf{q})$, which have been obtained by subtracting the local charge ($r$=c, orange curve), particle-particle ($r$=pp, green curve) or both ($r$=cpp, blue curve) correlation functions from $\Gamma_\text{sp}^{\nu\nu'\omega}$ in the BS Eq.~(\ref{equ:BS}), are larger than the corresponding lattice susceptibility of DMFT (red curve). The highest value is obtained when both local charge {\em and} particle-particle fluctuations are removed from $\Gamma_\text{sp}^{\nu\nu'\omega}$ (blue curve). 
Again, we observe that  the difference between the four curves is largest at the R point ($\mathbf{q}=\mathbf{\Pi}$).

Let us now turn our attention to the curves where the local spin susceptibility (and possibly one or both of the susceptibilities in the two other channels) have been subtracted from $\Gamma_\text{sp}^{\nu\nu'\omega}$ (black, brown, cyan, and pink lines in Fig.~\ref{fig:spinsuscq_comp}). Again, we observe the same hierarchy as for $T_N$ in Fig.~\ref{fig:compfluc}: For the considered $U$ values, all these curves are located below the corresponding DMFT results (red). At the lowest $U=1.0$, which is in the metallic regime of the Hubbard model, the differences are largest at the R point and almost vanish at the $\Gamma$ point, consistent with the discussion for weak coupling. However, at larger values of $U$ ($U=2.0$ and $U=4.0$) the suppression of $\chi_\text{sp}(\omega\!=\!0,\mathbf{q})$ due to the removal of local spin fluctuations from $\Gamma_\text{sp}^{\nu\nu'\omega}$ occurs almost equally at all momenta $\mathbf{q}$ and eventually leads to a complete vanishing of the lattice susceptibility at $U=4.0$. This behavior can be understood by the fact that $U=2.0$ and $U=4.0$ are located in the crossover/local moment regime of the Hubbard model where the suppression of the bubble $\chi_{0,\mathbf{q}}^{\nu\nu'\omega}$ [see definition below Eq.~(\ref{equ:BS})] due to a diverging local self-energy has to be compensated by a correspondingly large vertex $\Gamma_\text{sp}^{\nu\nu'\omega}$ in the BS Eq.~(\ref{equ:BS}) in order to obtain a finite lattice spin susceptibility $\chi_\text{sp}(\omega,\mathbf{q})$. Such large value of $\Gamma_\text{sp}^{\nu\nu'\omega}$ is mainly governed by the large value of $\chi_\text{sp}^{\text{loc}}(\omega)$ and, hence, removing this contribution from $\Gamma_\text{sp}^{\nu\nu'\omega}$ results in a very small value or even a vanishing of $\chi_\text{sp}(\omega,\mathbf{q})$.


\section{Conclusions and Outlook}
\label{sec:conclusions}

We have introduced a fluctuation analysis of the two-particle generalized and physical susceptibilities to analyze second-order phase transitions and, in particular, the transition temperatures to corresponding ordered states. Our approach is based on a diagrammatic decomposition of the input quantities for the equation used to compute the target objects, i.e., the susceptibilities. More specifically, this input is the irreducible vertex $\Gamma$ in the channel of interest, which is the central ingredient in the Bethe-Salpeter equation from which the two-particle correlation functions are obtained. The addition/removal of specific diagrammatic contributions to $\Gamma$ then provides crucial information {\em how} these terms affect the related susceptibilities and derived quantities such as transition temperatures.

In this paper, we have exploited this idea to analyze the transition to the antiferromagnetically ordered state in the half-filled Hubbard model within DMFT. In this case, the irreducible spin vertex $\Gamma_\text{sp}^{\nu\nu'\omega}$ is purely local, i.e., it depends only on the frequencies. As a starting point, we have approximated this correlation function using a weak-coupling expansion that has allowed us to express the vertex in terms of the local charge, spin, and particle-particle susceptibilities. Switching these terms sequentially on and off makes it possible to identify the impact of each of these three types of fluctuations on $T_N$. We found that local spin fluctuations enhance the antiferromagnetic spin susceptibility and $T_N$ while local charge and particle-particle fluctuations tend to suppress them. At weak coupling, a suppression due to local particle-particle fluctuations prevails while in the intermediate coupling regime the local spin fluctuations start to dominate and lead to a ``negative'' screening (i.e., an enhancement) of $\Gamma_\text{sp}$ with respect to the bare $U$. At the same time the inclusion of the DMFT self-energy in the single-particle Green's function leads to a reduction of the bubble term in the Bethe-Salpeter equation and, hence, to a reduction of the AF spin susceptibility. Therefore, in this regime the formation of the local moment described by DMFT plays a Janus-faced role: At the one-particle level it leads to an increase of the self-energy and, hence, to a suppression of spectral weight at the Fermi level, which, in turn, reduces the bubble term [given below Eq.~(\ref{equ:BS})] and, consequently, $T_N$. At the two-particle level, on the other hand, the local moment is reflected in an {\em enhancement} of the local irreducible vertex via local spin fluctuations which, tends to enhance $T_N$. The interplay between these two opposing effects controls the actual value of $T_N$ within DMFT. 

At strong-coupling the perturbative expansion for $\Gamma_{\text{sp}}$ breaks down. We notice that in Ref.~\onlinecite{Harkov2021}, this problem was solved within the dual fermion framework by adding triangular vertices, i.e., spin-fermion coupling contributions, to the approximation. This would obviously also be possible for $\Gamma_\text{sp}$, which can lead to quantitative changes and a corresponding improvement of our results. While we leave this interesting question for future research work we have verified our findings at weak-to-intermediate coupling on a qualitative level following another route: We have performed an inverse fluctuation analysis, which consists in subtracting the different local fluctuations from the exact $\Gamma_\text{sp}$ of DMFT leding to the same conclusions as the perturbative approximation of the vertex function. 

To gain more insights into the phase transition to the antiferromagnetically ordered state at large values of $U$, we have put forward a strong coupling approximation for the calculation of $T_N$ where we have replaced the local DMFT generalized susceptibility by the one of the atomic limit. How this approximation relates to a corresponding one within the DF framework (see Ref.~\onlinecite{Krien2019a}) is an interesting question for future research work. We believe that such approximation could be useful for practical purposes when using diagrammatic methods\cite{Rohringer2018a}, e.g., to study regimes where fermions form bound pairs close to Bose-Einstein condensation\cite{DelRe2019}. 

In the last part of this paper, we have discussed the momentum dependence of the lattice spin susceptibility as obtained from the different ways how local charge, spin, and particle-particle fluctuations are included in $\Gamma_\text{sp}$. Our findings indicate that the purely local modification of this vertex function gives rise to a nonlocal change of the lattice spin susceptibility where the difference between the various approximations is largest at the antiferromagnetic wave vector $\mathbf{q}\!=\!\mathbf{\Pi}$ and becomes gradually smaller far away from this lattice point.

Finally, we want to stress that the presented method is not restricted to the case of DMFT and the antiferromagnetic phase transition of the Hubbard model but can be applied within any theory where explicit expressions for $\Gamma$ are available, such as the diagrammatic extensions of DMFT\cite{Rohringer2018a,Toschi2007a,Rubtsov2009,Rohringer2011,Rubtsov2012,Rohringer2013,Rohringer2016,Ayral2016,Ayral2016a,delre2020dynamical}, and for any collective mode such as, e.g., $d$-wave superconductivity in the two-dimensional Hubbard model\cite{Kitatani2019}.

\paragraph*{Acknowledgements}
We thank V. Harkov, F. Krien, A. N. Rubtsov and A. Toschi for useful discussions. We acknowledge financial support from the Deutsche Forschungsgemeinschaft  (DFG)  through Project No.~407372336 (G.R.) and from the U.S. Department of Energy, Office of Science,
Basic Energy Sciences, Division of Materials Sciences and Engineering under Grant No. DE-SC0019469 (L.D.R.). The work was supported by the North-German Supercomputing Alliance (HLRN).

\appendix
\section{Comparison between ED and QMC}
\label{app:compQMC}
\begin{figure}[t!]
    \centering
    \includegraphics[width=1.0\columnwidth]{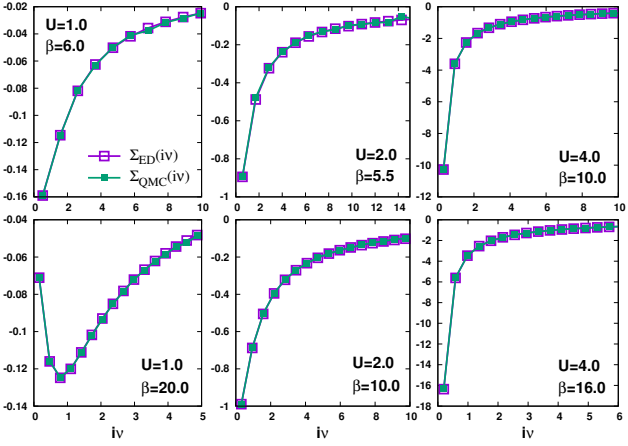}
    \caption{Imaginary part of the local self-energy $\Sigma(\nu)$ as a function of the fermionic Matsubara frequency $\nu$ calculated with ED (violet empty squares) and QMC (green filled squares) for three different values of $U$ at weak (left panels), intermediate (middle panels), and strong (right panels) coupling at high temperatures (upper panels) and low temperatures (lower panels).}
    \label{fig:CompareSigma}
\end{figure}

\begin{figure}[t!]
    \centering
    \includegraphics[width=1.0\columnwidth]{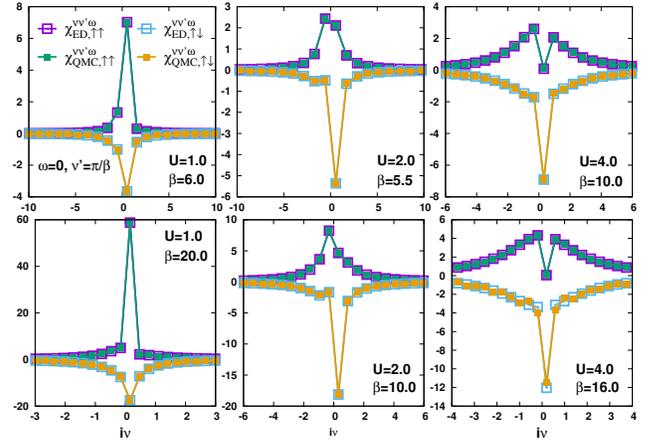}
    \caption{Real part of the local generalized susceptibilities $\chi_{\sigma\sigma'}^{\nu(\nu'=\pi/\beta)(\omega=0)}$ as a function of $\nu$ for fixed $\nu'=\frac{\pi}{\beta}$ and $\omega=0$ for the same values of $U$ and $\beta$ as in Fig.~\ref{fig:CompareSigma}. The two spin projections $\uparrow\uparrow$ and $\downarrow\downarrow$ are indicated with different colors.}
    \label{fig:CompareChi}
\end{figure}

\begin{figure}[t!]
    \centering
    \includegraphics[width=1.0\columnwidth]{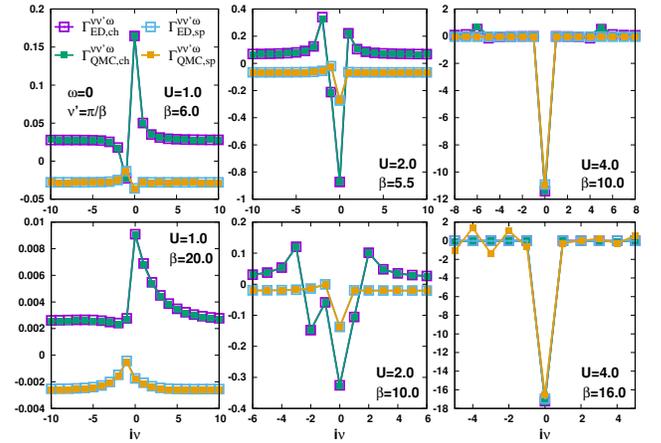}
    \caption{Real part of the local irreducible vertex $\Gamma_{\text{ch}/\text{sp}}^{\nu(\nu'=\pi/\beta)(\omega=0)}$ as a function of $\nu$ for fixed $\nu'=\frac{\pi}{\beta}$ and $\omega=0$ for the same values of $U$ and $\beta$ as in Fig.~\ref{fig:CompareSigma}. The charge and spin channel are depicted in different colors.}
    \label{fig:CompareGamma}
\end{figure}

\begin{figure}[t!]
    \centering
    \includegraphics[width=1.0\columnwidth]{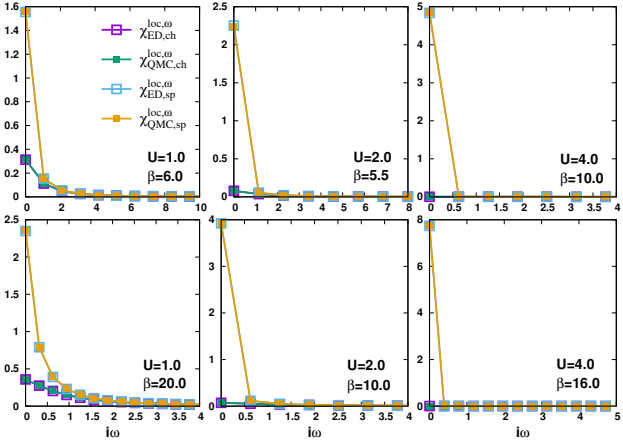}
    \caption{Physical susceptibilities $\chi^\text{loc}_{\text{ch}/\text{sp}}(\omega)$ as a functions of the bosonic frequency $\omega$ for the same values of $U$ and $\beta$ as in Fig.~\ref{fig:CompareSigma}. The charge and spin channel are depicted in different colors.}
    \label{fig:CompareChiphys}
\end{figure}

\begin{figure}[t!]
    \centering
    \includegraphics[width=1.0\columnwidth]{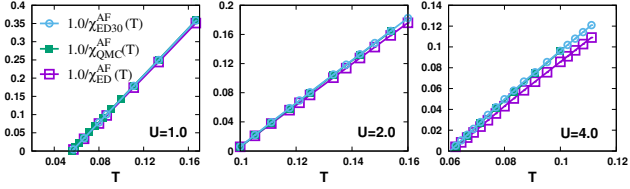}
    \caption{Inverse antiferromagnetic susceptibility $\chi_\text{sp}(\omega=0,\mathbf{q}=\boldsymbol{\Pi})$ as a function of temperature for the same values of $U$ as in Fig.~\ref{fig:CompareSigma}. $\chi_{\text{ED}30}^\text{AF}(T)$ corresponds to ED results where 30 positive and negative Matsubara frequencies have been used (instead of 160), which is the same as for the QMC calculations.}
    \label{fig:CompareChiAF}
\end{figure}

In this section, we benchmark our ED\cite{Georges1996,capone2007} results for the local and nonlocal correlation functions with corresponding QMC\cite{Gull2011a} data. As QMC solver the CT-HYB implementation of the w2dynamics\cite{Wallerberger2019} package has been used. We present our comparison for all relevant local correlation functions for three representative values of $U$ at weak ($U=1.0$), intermediate ($U=2.0$), and strong ($U=4.0$) coupling (corresponding to the left, middle, and right panels in Figs.  \ref{fig:CompareSigma}-\ref{fig:CompareChiphys}, respectively). The frequency-dependent local correlation functions in Figs.~\ref{fig:CompareSigma}-\ref{fig:CompareChiphys} are depicted for a high temperature (upper panels) and a lower temperature closer to the phase transition (lower panels) as a function of one Matsubara frequency. In particular, the generalized susceptibilities $\chi_{\sigma\sigma'}^{\nu\nu'\omega}$ in Fig.~\ref{fig:CompareChi} and the irreducible vertices $\Gamma_{\text{ch}/\text{sp}}^{\nu\nu'\omega}$ in Fig.~\ref{fig:CompareGamma} are shown as a function of the fermionic Matsubara frequency $\nu$ for a fixed $\nu'=\frac{\pi}{\beta}$ and $\omega=0$. The inverse antiferromagnetic susceptibility $\chi_\text{AF}^{-1}(T)$ is plotted in Fig.~\ref{fig:CompareChiAF} as a function of temperature. Let us point out that the half-filled Hubbard model on a bipartite lattice (such as the simple cubic lattice in three dimensions) is particle-hole symmetric, which is also true for the related AIM. This implies that the local self-energy is purely imaginary (apart from the real constant Hartree term $\frac{Un}{2}$) while the generalized susceptibilities and the irreducible vertex in Figs.~\ref{fig:CompareChi} and \ref{fig:CompareGamma} are purely real (the local physical susceptibilities in Fig.~\ref{fig:CompareChiphys} are always real quantities, also away from particle-hole symmetry).

The data point given by $U\!=\!1.0$ and $\beta=20.0$ (left lower panels) corresponds to the metallic regime of the DMFT phase diagram, which is indicated by the non-monotonous behavior of the (imaginary part of the) self-energy at low frequencies (see left-lower panel in Fig.~\ref{fig:CompareSigma}). At $U\!=\!4.0$ on the other hand (right panels in Fig.~\ref{fig:CompareSigma}), $\Sigma(\nu)$ features an insulating behavior, which can be inferred from its monotonous behavior and its large values at low frequencies. The remaining three data points [($U\!=\!1.0$, $\beta\!=\!6.0$), ($U\!=\!2.0$, $\beta\!=\!5.5$), ($U\!=\!2.0$, $\beta\!=\!10.0$)] belong to the so-called crossover region, which is located between the metallic and the insulating phase of the Hubbard model at high temperatures (see, for instance, Ref.~\onlinecite{Rohringer2011}). In this regime, the self-energy already shows an insulating-like monotonous behavior but its size is still moderate.

In general one observes an excellent agreement for all local one- and two-particle correlation functions obtained from ED with the corresponding QMC results. Remarkably, this even holds for $\Gamma_{\text{ch}/\text{sp}}^{\nu\nu'\omega}$, which is obtained from $\chi_{\sigma\sigma'}^{\nu\nu'\omega}$ via a matrix inversion [see Eq.~(\ref{equ:BS})]. Only at large frequencies small fluctuations can be observed in the QMC data, which originate from the statistical noise of the QMC calculation.

For the inverse of the antiferromagnetic susceptibility in Fig.~\ref{fig:CompareChiAF} some deviations can be observed between ED (empty violet squares) and QMC (filled green squares), in particular at strong coupling. These differences, however, originate from the different number of frequencies, which have been used for performing the sum over the generalized susceptibility in Eq.~(\ref{equ:sumgeneralizedsusc}) to obtain the nonlocal spin susceptibility [230 frequencies for ED vs 60 frequencies for QMC]. Reducing the number of frequencies also in the ED calculation (see empty blue circles in Fig.~\ref{fig:CompareChiAF}) restores the excellent agreement between the ED and the QMC results. 

\section{Strong coupling formula}
\label{app:alanalytic}


In this section, we provide some details regarding the calculation of $T_N$ in the strong-coupling regime [Eq.~(\ref{equ:alanalytic})] where the local generalized susceptibility $\bar{\bar{\chi}}_\text{sp}(\omega\!=\!0)$ in Eq.~(\ref{eq:bse:bethe}) has been replaced by the one of the atomic limit $\bar{\bar{\chi}}_\text{sp}^\text{AL}(\omega\!=\!0)$. More specifically, the inverse of this matrix in the fermionic $\nu$-$\nu'$ frequency space is required in Eq.~(\ref{eq:bse:bethe}), which has been obtained analytically (for the AL at half filling) in Ref.~\onlinecite{Thunstr2018}. For $\omega\!=\!0$ it reads [consider Eq.~(19) in Ref.~\onlinecite{Thunstr2018} in the spin channel]:
\begin{align}
\label{equ:chiALinverse}
\left[\bar{\bar{\chi}}_\text{sp}^\text{AL}(\omega=0)\right]^{-1}_{\nu\nu'}=&a_{0}^\nu\left[\delta_{\nu\nu'}-\delta_{\nu(-\nu')}\right]\nonumber\\+&b_{0}^{\nu}\left[\delta_{\nu\nu'}+\delta_{\nu(-\nu')}\right]+\sum_{i=1}^2b_{i}^{\nu}b_{i}^{\nu'},
\end{align}
where the quantities $a_0^\nu$ and $b_i^\nu$ ($i\!=\!0,1,2$) are given by
\begin{subequations}
\label{equ:defa0bi}
\begin{align}
\label{equ:defa0b0} &a_0^\nu=\frac{1}{2\beta}\left(\nu^2+\frac{U^2}{4}\right)&b_0^\nu=\frac{1}{2\beta}\frac{\left(\nu^2+\frac{U^2}{4}\right)^2}{\nu^2+B^2}\\
\label{equ:defb1b2} &b_1^\nu=\frac{\sqrt{C}}{\beta}\frac{1}{\nu^2+B^2} & b_2^\nu=b_2=\frac{\sqrt{-U}}{\beta} ,
\end{align}
\end{subequations}
and the frequency-independent quantities $B$ and $C$ (which, however, depend on $\beta$ and $U$) are defined as
\begin{align}
\label{equ:defBC}
B=\frac{U}{2}\sqrt{\frac{e^{\frac{\beta U}{2}}-3}{e^{\frac{\beta U}{2}}+1}}, \quad
C=\frac{U^5}{16}\frac{\left(1-\frac{4B^2}{U^2}\right)^2}{1-\frac{U\tanh\left(\frac{\beta}{2}B\right)}{2B}}.
\end{align}
Since we are interested in the strong-coupling regime at intermediate to low temperatures, we can assume that $\frac{\beta U}{2}\!>\!\log 3$ and, hence, $B$ and $C$ are positive real numbers. According to Eq.~(\ref{eq:bse:bethe}), we have to subtract $\frac{t^2}{\beta}\delta_{\nu\nu'}$ from $\left[\bar{\bar{\chi}}_\text{sp}(\omega=0)\right]^{-1}_{\nu\nu'}$. This subtraction modifies only the diagonal contributions $a_0^\nu$ and $b_0^\nu$ of $\left[\bar{\bar{\chi}}_\text{sp}^\text{AL}(\omega=0)\right]^{-1}_{\nu\nu'}$ which, hence, become
\begin{subequations}
\label{equ:a0tb0t}
\begin{align}
\label{equ:a0t} & a_0^\nu\rightarrow a_0^\nu-\frac{t^2}{2\beta}=a_t^\nu=\frac{1}{2\beta}\left(\nu^2+\frac{U^2}{4}-t^2\right)\\
\label{equ:b0t} & b_0^\nu\rightarrow b_0^\nu-\frac{t^2}{2\beta}=b_t^\nu=\frac{1}{2\beta}\left[\frac{\left(\nu^2+\frac{U^2}{4}\right)^2}{\nu^2+B^2}-t^2\right].
\end{align}
\end{subequations}
To calculate the generalized antiferromagnetic susceptibility $\bar{\bar{\chi}}_\text{sp}(0,\boldsymbol{\Pi})$ in Eq.~(\ref{eq:bse:bethe}) we have to invert the matrix on the right hand side of this equation. This matrix is exactly the one given in Eq.~(\ref{equ:chiALinverse}) but with $a_0^\nu$ and $b_0^\nu$ replaced by $a_t^\nu$ and $b_t^\nu$, respectively. The inversion of this matrix can be performed analytically by means of the Sherman-Morrison-Woodbury matrix identity\cite{Hager1989}, completely analogous to the procedure outlined in Ref.~\onlinecite{Thunstr2018}. This yields
\begin{align}
\label{equ:chigeneralaffinal}
\left[\bar{\bar{\chi}}_\text{sp}(0,\boldsymbol{\Pi})\right]_{\nu\nu'}=&\frac{1}{4a_{t}^{\nu}}[\delta_{\nu\nu'}-\delta_{\nu(-\nu')}]+\frac{1}{4b_{t}^{\nu}}[\delta_{\nu\nu'}+\delta_{\nu(-\nu')}] \nonumber\\
&-\frac{1}{4b_{t}^{\nu}b_{t}^{\nu'}}\sum_{k,l=1}^2b_{k}^{\nu}(M^{-1})_{kl}b_{l}^{\nu'},
\end{align}
where the $2\times 2$ matrix $M$ is given by
\begin{align}
\label{equ:defM}
M_{kl} = \delta_{kl} + \sum_\nu\frac{b_{k}^{\nu}b_{l}^{\nu}}{2b_{t}^{\nu}},\quad k,l=1,2.
\end{align}
To obtain the AF susceptibility, we have to sum Eq.~(\ref{equ:chigeneralaffinal}) over the fermionic Matsubara frequencies $\nu$ and $\nu'$, which yields
\begin{align}
\label{equ:chiaffinal}
\chi_\text{AF}=\frac{1}{\beta^2}\sum_\nu\frac{1}{2b_t^\nu}-\left(\frac{1}{\beta}\sum_\nu\frac{b_k^\nu}{2b_t^\nu}\right)(M^{-1})_{kl}\left(\frac{1}{\beta}\sum_\nu\frac{b_l^\nu}{2b_t^\nu}\right).
\end{align}
The transition to the AF is signaled by a divergence of $\chi_\text{AF}$. In Eq.~(\ref{equ:chiaffinal}), such a divergence can arise in two ways: (i) the function $b_t^\nu$ can diverge for a given frequency $\nu$. However, as it has been shown in Ref.~\onlinecite{Thunstr2018}, such singularity would appear in both terms on the right hand side of Eq.~(\ref{equ:chiaffinal}) and cancel. (ii) The only other possibility for $\chi_\text{AF}$ to diverge consists in a singularity of the $2\times 2$ matrix $M$. Hence, $T_N$ can be determined by the condition that the determinant of $M$ vanishes. Consequently, the remaining task is to evaluate the Matsubara sums in Eq.~(\ref{equ:defM}) for the three different matrix elements of the symmetric $2\times 2$ matrix $M$:
\begin{widetext}
\begin{subequations}
\label{equ:matsums}
\begin{align}
\label{equ:matsums11}M_{11}=&1+\sum_\nu\frac{b_1^\nu b_1^\nu}{2b_t^\nu}=1+\frac{C}{\beta}\sum_\nu\frac{1}{\left(\nu^2+B^2\right)^2}\frac{1}{\frac{\left(\nu^2+\frac{U^2}{4}\right)^2}{\nu^2+B^2}-t^2}=1+\frac{C}{\beta}\sum_\nu\frac{1}{\nu^2+B^2}\frac{1}{\nu^2+R_+^2}\frac{1}{\nu^2+R_-^2}\nonumber\\=&1+C\left[\frac{\tanh\left(\frac{\beta}{2}B\right)}{2B(R_+^2-B^2)(R_-^2-B^2)}+\frac{\tanh\left(\frac{\beta}{2}R_+\right)}{2R_+(R_+^2-R_-^2)(R_+^2-B^2)}+\frac{\tanh\left(\frac{\beta}{2}R_-\right)}{2R_-(R_-^2-R_+^2)(R_-^2-B^2)}\right]\\
\label{equ:matsums12}M_{12}=&M_{21}=\sum_\nu\frac{b_1^\nu b_2^\nu}{2b_t^\nu}=\frac{\sqrt{-UC}}{\beta}\sum_\nu\frac{1}{\nu^2+B^2}\frac{1}{\frac{\left(\nu^2+\frac{U^2}{4}\right)^2}{\nu^2+B^2}-t^2}=\frac{\sqrt{-UC}}{\beta}\sum_\nu\frac{1}{\nu^2+R_+^2}\frac{1}{\nu^2+R_-^2}\nonumber\\=&\sqrt{-UC}\left[\frac{\tanh\left(\frac{\beta}{2}R_+\right)}{2R_+(R_+^2-R_-^2)}+\frac{\tanh\left(\frac{\beta}{2}R_-\right)}{2R_-(R_-^2-R_+^2)}\right]\\
\label{equ:matsums22}M_{22}=&1+\sum_\nu\frac{b_2^\nu b_2^\nu}{2b_t^\nu}=1-\frac{U}{\beta}\sum_\nu\frac{1}{\frac{\left(\nu^2+\frac{U^2}{4}\right)^2}{\nu^2+B^2}-t^2}=1-\frac{U}{\beta}\sum_\nu\frac{\nu^2+B^2}{\left(\nu^2+R_+^2\right)\left(\nu^2+R_-^2\right)}\nonumber\\=&1-U\left[\frac{\tanh\left(\frac{\beta}{2}R_+\right)}{2R_+}\frac{R_+^2-B^2}{R_+^2-R_-^2}+\frac{\tanh\left(\frac{\beta}{2}R_-\right)}{2R_-}\frac{R_-^2-B^2}{R_-^2-R_+^2}\right],
\end{align}
\end{subequations}
\end{widetext}
where
\begin{subequations}
\label{equ:rplusminus}
\begin{align}
\label{equ:rplus} &R_+=\sqrt{\frac{U^2}{4}-\frac{t^2}{2}-\frac{t}{2}\sqrt{t^2-U^2+4B^2}}\\
\label{equ:rminus} &R_-=\sqrt{\frac{U^2}{4}-\frac{t^2}{2}+\frac{t}{2}\sqrt{t^2-U^2+4B^2}}
\end{align}
\end{subequations}
are real positive parameters if $U\!>\!2t$ (which is fulfilled in the strong-coupling limit). In Eqs.~(\ref{equ:matsums}), partial fraction decomposition has been extensively used, which reduced the various summations over the fermionic Matsubara frequency $\nu$ to the standard Matsubara sum
\begin{equation}
\label{equ:standarmatsum}
\frac{1}{\beta}\sum_\nu\frac{1}{\nu^2+X^2}=\frac{\tanh\left(\frac{\beta}{2}X\right)}{2X},
\end{equation}
for a positive real number $X$.

As already mentioned, the condition $M_{11}M_{22}-M_{12}^2\!=\!0$ corresponds to the divergence of $\chi_\text{AF}$ and---considering Eqs.~(\ref{equ:matsums})---represents, hence, a very complicated transcendental equation for the determination of the transition temperature $T_N$. Even a numerical solution turns out to be difficult due to the cancellation of several exponentially suppressed terms. For instance, considering the constant $C$ [see Eq.~(\ref{equ:defBC})] in the limit $\frac{\beta U}{2}\!\rightarrow\!\infty$, both the numerator and the denominator decay exponentially as $e^{-\beta U}$ (with logarithmic corrections, i.e., with correction terms of the order $\log e^{-\frac{\beta U}{2}}\!=\!-\frac{\beta U}{2}$). We have hence expanded all expressions in Eqs.~(\ref{equ:matsums}) in terms of
\begin{equation}
\label{equ:expand}
y=e^{-\frac{\beta U}{2}}\quad\Leftrightarrow\quad\frac{\beta U}{2}=-\log y.
\end{equation}
Neglecting all terms of the form $e^{-\sqrt{1-\frac{4t^2}{U^2}}\frac{\beta U}{2}}$ and of higher orders in $e^{-\frac{\beta U}{2}}$ then gives for the determinant of $M$:
\begin{align}
\label{equ:detfinal}
\det M=\left(1-\frac{1}{\sqrt{1-\frac{4t^2}{U^2}}}\right)\left(1-\frac{U^2}{t^2}\frac{1}{\beta U-1}\right).
\end{align}
Since the first factor on the right-hand side of this equation is always smaller than $0$, the vanishing of the determinant corresponds to the second term being $0$, i.e., $(1-\frac{U^2}{t^2}\frac{1}{\beta U-1})\!=\!0$. Solving this equation for $T\!=\!\frac{1}{\beta}$ leads to Eq.~(\ref{equ:alanalytic}) in the main text.


%

\end{document}